\documentclass[final,3p,11pt]{elsarticle}

\usepackage{amssymb}
\usepackage{graphicx}
\usepackage{caption2}
\usepackage{epsfig}
\usepackage{amsmath}
\usepackage{amsfonts}

\newtheorem{Theorem}{\hspace{-0.001in}{\bf Theorem}}
\newtheorem{Conjecture}{\hspace{-0.001in}{\em Conjecture}}

\newtheorem{Lemma}{\hspace{-0.001in}{\bf Lemma}}

\def\done{\hspace*{\fill} \rule{1.8mm}{2.5mm}}
\newtheorem{Definition}{\hspace{-0.001in}{\bf Definition}}

\captionstyle{center}

\journal{Computer Networks}

\begin{document}

\begin{frontmatter}

\title{\textbf{On Oligopoly Spectrum Allocation Game in Cognitive Radio Networks with Capacity
Constraints}}

\author[xyd]{Yuedong Xu}
\ead{ydxu@cse.cuhk.edu.hk}
\author[xyd]{John C.S. Lui\corref{cor1}}
\ead{cslui@cse.cuhk.edu.hk}
\author[els]{Dah-Ming Chiu}
\ead{dmchiu@ie.cuhk.edu.hk}

\cortext[cor1]{Corresponding author}

\address[xyd]{Department of Computer Science \& Engineering, The Chinese University of Hong Kong, Hong Kong}
\address[els]{Department of Information Engineering, The Chinese University of Hong Kong, Hong Kong}

\begin{abstract}
\label{section:abstract}

Dynamic spectrum sharing is a promising technology to improve
spectrum utilization in the future wireless networks. The flexible
spectrum management provides new opportunities for licensed primary
user and unlicensed secondary users to reallocate the spectrum
resource efficiently. In this paper, we present an oligopoly pricing
framework for dynamic spectrum allocation in which the primary users
sell excessive spectrum to the secondary users for monetary return.
We present two approaches, the \emph{strict constraints (type-I)}
and the \emph{QoS penalty (type-II)}, to model the realistic
situation that the primary users have limited capacities. In the
oligopoly model with strict constraints, we propose a low-complexity
searching method to obtain the Nash Equilibrium and prove its
uniqueness. When reduced to a duopoly game, we analytically show the
interesting gaps in the leader-follower pricing strategy. In the QoS
penalty based oligopoly model, a novel variable transformation
method is developed to derive the unique Nash Equilibrium. When the
market information is limited, we provide three myopically optimal
algorithms ``StrictBEST'', ``StrictBR'' and ``QoSBEST'' that enable
price adjustment for duopoly primary users based on the Best
Response Function (BRF) and the bounded rationality (BR) principles.
Numerical results validate the effectiveness of our analysis and
demonstrate the fast convergence of ``StrictBEST'' as well as
``QoSBEST'' to the Nash Equilibrium. For the ``StrictBR'' algorithm,
we reveal the chaotic behaviors of dynamic price adaptation in
response to the learning rates.

\end{abstract}


\end{frontmatter}

\section{{\bf Introduction}}
\label{section:intro}

Wireless spectrum has become the scarcest resource due to the
dramatic development of mobile telecommunication industry in the
last decades. However, recent studies by the Federal Communications
Commission (FCC) show that traditional fixed allocation policy
results in very low efficiency in radio spectrum utilization. The
increasing spectrum demand, together with the resource scarcity,
gives rise to the development of \emph{cognitive radio networks}
that enable dynamic spectrum access. Within a dynamic spectrum
access system, radio spectrum resources are allocated by agile
management schemes in terms of \emph{spectrum market} among the
unlicensed (i.e., secondary) users and the licensed (i.e., primary)
users \cite{Mobihoc08:Zhang}. When the possessed spectrum is not
fully utilized, a primary user has an opportunity to sell the
excessive spectrum to the secondary users for monetary payoff. This
is also referred to as spectrum trading mechanism in
\cite{JSAC08:Niyato} which involves spectrum selling and purchasing
processes. Therefore, it is natural to consider the spectrum
allocation in the perspective of economic models and market
strategies.

In such an emerging network scenario, multiple primary users coexist
in the same geographical site and compete for the access or the
purchase of secondary users equipped with cognitive radios. Hence,
an important problem for the spectrum trading is how the primary
users set prices of per-unit spectrum in a competitive market. For
example, if a primary user sets a very low price, it might result in
the loss of revenue (or profit). On the contrary, if the price is
set too expensive, the secondary users are inclined to purchase from
other spectrum holders. Niyato et.al \cite{JSAC08:Niyato} initially
introduce the oligopoly pricing theory to characterize the
interactions between the spectrum abundant side (primary users) and
the demanding side (secondary users). In the oligopoly spectrum
market, a commonly used quadratic utility is adopted to quantify the
spectrum demand of the secondary service, and each primary user aims
to maximize the individual profit. In \cite{Mobihoc08:Zhang}, Jia
and Zhang study the competitions and dynamics of spectrum allocation
in a duopoly market via a non-cooperative two-stage game. However,
authors in \cite{JSAC08:Niyato} do not consider an important feature
that a primary user usually has limited capacity to lease. Authors
in \cite{Mobihoc08:Zhang} mainly focus on the situation that both
two wireless service providers have limited spectrum capacities in
the price competition stage. In fact, the price competition in a
more general oligopoly game is rather difficult to be analyzed when
the constraints of spectrum capacity are incorporated.

In this paper, we investigate the competitive pricing of a general
oligopoly spectrum game. Distinguished from previous work, our study
concentrates on the capacity-constrained pricing that is quite
common to the primary users. To characterize the limitation in
leasing spectrum resource, we employ two approaches: the
\emph{strict constraints} (type-I) and the \emph{QoS penalty
functions} (type-II). In the market model with type-I constraints,
each primary user has a certain spectrum bound so that it might not
be able to provide the best spectrum demand. We address the
following challenging issues: a) is there a unique Nash Equilibrium
(\emph{NE}) in such a capacity-constrained spectrum game? b) if YES,
how to find the \emph{NE} efficiently? c) especially in duopoly
games, what are the impacts of capacity constrains on the \emph{NE}s
and the system dynamics? In the type-II market model, the capacity
constraint of a primary user is absorbed in the utility as a barrier
penalty function. This corresponds to the scenario that the leased
spectrum is transferred from the existing services of primary users.
They offer spectrum for monetary return, however, at the cost of QoS
decrease of primary services. Here, we model the QoS of a primary
service as a function of the queueing delay. Generally, explicit
solution does not exist in such an oligopoly spectrum allocation
game. We present a novel method to discover the \emph{NE} and to
prove its uniqueness. Consider the fact that a primary user usually
has no knowledge of the utilities and the price-demand functions of
its opponents, we develop a set of price adjustment algorithms based
on the best response dynamics and the bounded rationality
principles.

To summarize, our contributions are:

\begin{itemize}

\item We formulate two oligopoly market models to characterize the
capacity limitations: the strict constraints (type-I) and the QoS
based penalty functions (type-II). Given the above market models,
the primary users compete for revenue or utility maximization by
deciding the prices of per-unit spectrum.

\item In the type-I model, we propose a novel searching method to
find a Nash Equilibrium and prove its uniqueness. Interestingly, we
find the revenue gaps in the duopoly Stackelberg game with type-I
constraints.

\item We present two algorithms,
\emph{StrictBEST} and \emph{StrictBR}, to adjust prices dynamically
based on the best response dynamics and the bounded rationality when
the market information is limited.

\item In the type-II model, we present an interesting variable
transformation method to derive the Nash Equilibrium and prove its
uniqueness. The \emph{QoSBEST} algorithm is proposed to perform
spectrum pricing based on the best response dynamics.

\item We demonstrate the nonlinear dynamic behaviors in the
\emph{StrictBR} algorithm when the learning rates vary.

\end{itemize}

The rest of this paper is organized as follows. In section
\ref{section:model}, we present the system models of
capacity-constrained spectrum market. In section \ref{section:type1}
and \ref{section:type2}, we analyze the \emph{NE}s of the
noncooperative oligopoly market with type-I and type-II constrains
respectively. Section \ref{section:evaluation} evaluates the
analysis and performance of proposed schemes. We present an overview
of related work in section \ref{section:related} and conclude in
section \ref{section:conclusion}.

\section{{\bf System Model}}
\label{section:model} In this section, we present mathematical
models to characterize the dynamic spectrum allocation in cognitive
radio networks. To capture the realistic spectrum market, two types
of capacity constraints are incorporated.

\subsection{Agile Spectrum Market}

\begin{figure}[!htb]
\centering
\includegraphics[width = 3in, height=3in]{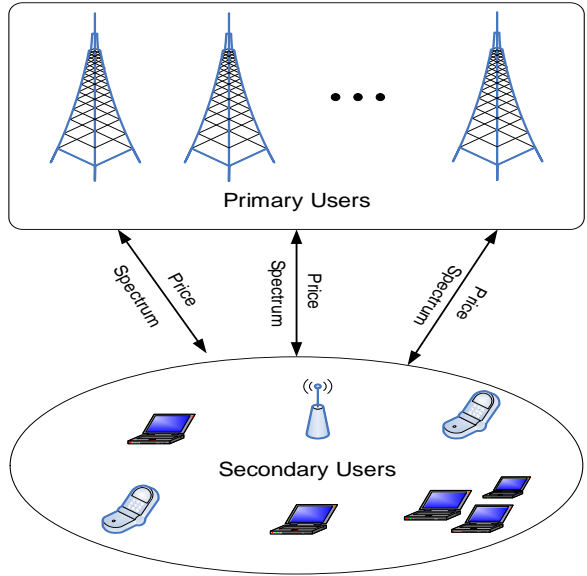}
\caption{\small{Spectrum Market Structure}}
\label{figure:fig2.1.1_structure}
\end{figure}

We consider the cognitive radio network where multiple primary users
(PUs) or wireless service providers (WSPs) compete for a shared pool
of secondary users. The secondary users are the static/mobile
devices equipped with cognitive radio technologies. The primary
users are the infrastructure based wireless operators or the
licensed spectrum holders. They are usually treated as the spectrum
brokers that lease the unused frequency to the secondary users for
monetary payoff. We show the structure of a spectrum market in
Fig.\ref{figure:fig2.1.1_structure} with a number of primary users
and the common secondary users. In this spectrum market, the demands
of secondary users depend on the prices of per-unit spectrum. Each
primary user chooses its own price to compete for the secondary
users' subscription.

\subsection{Secondary Users}

We characterize the spectrum demands of secondary users in the
oligopoly market. Define $\mathcal{N}=\{1,2,\cdots, N\}$ as the set
of primary users. Let $p_i$ be the price of the $i^{th}$ primary
user. Denote $q_i$ to be the quantity of spectrum that secondary
users buy from the $i^{th}$ primary user. We define the utility of
an average secondary user as a quadratic and concave function
\cite{Book99:Vives}\cite{JSAC08:Niyato}:
\begin{equation}
u(\mathbf{q}) = \sum_{i=1}^{N}\alpha_iq_i -
\frac{1}{2}\big(\sum_{i=1}^{N}\beta_iq_i^2+2\mu\sum_{i=1}^{N}\sum_{j\neq
i}q_iq_j\big) -
\sum_{i=1}^{N}p_iq_i,\label{equation:sec2.1_userutility}
\end{equation}
\noindent where $\alpha_i$, $\beta_i$ are positive constants for all
$i\in\mathcal{N}$. Here, $\alpha_i$ denotes the spectral efficiency
of wireless transmission by a secondary user using the frequency
owned by the $i^{th}$ primary user \cite{JSAC08:Niyato}. The
spectral efficiencies of primary users can be the same, or
different, depending on their center frequencies. For instance, if
the center frequency of the $i^{th}$ primary user is high, secondary
users may experience large path loss (or low spectral efficiency
equivalently) when purchasing spectrum from this primary user.
Similar to previous work, we also take the spectrum substitutability
into account via the parameter $\mu$. If $\mu$ is 0, a secondary
user cannot switch among the primary users. When $0< \mu < \beta_i$,
a secondary user can switch among the primary users depending on the
spectral efficiency and the price of per-unit spectrum. For example,
if one primary user increases its price of per-unit of spectrum,
some of the secondary users may buy spectrum from other primary
users, and vice versa. When $\alpha_i = \alpha$ and $\beta_i = \mu$
for all $i\in\mathcal{N}$, the spectra of primary users are
perfectly substitutive for secondary users. Taking the first-order
derivative of the utility function with respect to $q_i$ and letting
it be 0, we obtain the purchase price of secondary users from the
$i^{th}$ primary user:
\begin{eqnarray}
p_i = \alpha_i-\beta_iq_i- \mu \sum_{j\neq i}q_j, \;\;\; \forall
i,j\in\mathcal{N}. \label{equation:sec2.2_pricefunc}
\end{eqnarray}
\noindent The concave function in
Eqn.(\ref{equation:sec2.1_userutility}) characterizes user
satisfaction in terms of the purchased spectrum. In order to
guarantee the concavity of utility function, its Jacobian matrix
should be negative definite, that is,
\begin{eqnarray}
\frac{\partial u(\mathbf{q})}{\partial \mathbf{q}} = - \mathbf{T} =
 - \left[
\begin{array}{cccc}
\beta_{1} & \mu & \ldots & \mu \\
\mu & \beta_{2} & \ldots & \mu \\
\vdots & \vdots & \ddots & \vdots \\
\mu & \mu & \ldots & \beta_N
\end{array} \right] < 0,
\end{eqnarray}
\noindent where $\mathbf{T}$ denotes the above matrix. Note that the
concavity of utility function is equivalent to the positive
definiteness of $\mathbf{T}$. Here, we present a necessary condition
in Lemma \ref{Lemma:PositiveDefinite} to set market parameters.
\begin{Lemma}
{\em The matrix $\mathbf{T}$ is positive definite if the market
parameters has $\beta_i > \mu >0$ for all $i \in \mathcal{N}$.}
\label{Lemma:PositiveDefinite}
\end{Lemma}
\textbf{Proof:} Please refer to the Appendix. \done

For the primary users, their strategies are to set prices of
per-unit spectrum in the oligopoly market. Thus, the demand function
can be expressed as the following:
\begin{eqnarray}
\mathbf{q} = \left[ \begin{array}{cccc}
\beta_{1} & \mu & \ldots & \mu \\
\mu & \beta_{2} & \ldots & \mu \\
\vdots & \vdots & \ddots & \vdots \\
\mu & \mu & \ldots & \beta_N
\end{array} \right]^{-1}\cdot
\left(\left[ \begin{array}{c}
\alpha_1 \\
\alpha_2 \\
\vdots \\
\alpha_N
\end{array} \right] - \mathbf{p}\right) \;
\label{equation:sec2.3_demandfuncmatrix}
\end{eqnarray}
\noindent where $\mathbf{p}$ is the price vector and $\mathbf{q}$ is
the spectrum demand vector. Given the price vector, the spectrum
demand of primary user $i$ is rewritten by:
\begin{eqnarray}
q_i = f_i(\mathbf{p}) = a_i - b_ip_i + \sum_{j\neq i}c_{ij}p_j,
\;\;\; \forall
i,j\in\mathcal{N},\label{equation:sec2.4_demandfunceqn}
\end{eqnarray}
\noindent where $b_i$ as well as $c_{ij}$ are variables computed
through Eqn.(\ref{equation:sec2.3_demandfuncmatrix}) and $a_i =
\alpha_i b_i - \sum_{j\neq i} c_{ij}\alpha_j$. Especially, $c_{ij} =
c_{ji}$ and $c_{ii}=-b_i$ due to the symmetry of convert matrix. The
market parameters transformed from
Eqn.(\ref{equation:sec2.3_demandfuncmatrix}) satisfy the following
property,
\begin{Lemma}
{\em The parameters that characterize demand-price function in
Eqn.(\ref{equation:sec2.4_demandfunceqn}), i.e.$\; b_i $ and
$c_{ij}$ $(i\neq j)$, are positive, given the conditions $\beta_i >
\mu > 0$ for all $i,j \in \mathcal{N}$.}
\label{Lemma:PositiveVariables}
\end{Lemma}
\textbf{Proof:} Please refer to the Appendix. \done

\noindent In the market model, if the spectral efficiencies (i.e.
$\alpha_i, \forall i\in\mathcal{N}$) are close to each other, one
can guarantee the positivity of $a_i$ by choosing $\beta_i$ and
$\mu$ appropriately. We give an example of spectrum market with only
two primary users. The utility function results in linear inverse
demand functions
\begin{eqnarray}
p_1 &=& \alpha_1 - \beta_1q_1 -\mu
q_2,\\
p_2 &=& \alpha_2 - \beta_2q_2 -\mu
q_1.\label{equation:sec2.5_linearinverse}
\end{eqnarray}
\noindent The relationship between prices and demands can be
represented by an alternative form:
\begin{eqnarray}
q_1 &=& f_1(\mathbf{p}) = a_1 - b_1p_1 + cp_2,
\label{equation:sec2.6_demandprice}\\
q_2 &=& f_2(\mathbf{p}) = a_2 - b_2p_2 +
cp_1.\label{equation:sec2.7_demandprice}
\end{eqnarray}
\noindent According to Eqn.(\ref{equation:sec2.3_demandfuncmatrix}),
$a_i$ and $b_i$ are calculated by:
$a_i=(\alpha_i\beta_j-\alpha_j\mu)/(\beta_1\beta_2-\mu^2)$,
$b_i=\beta_j/(\beta_1\beta_2-\mu^2)$ and
$c=\mu/(\beta_1\beta_2-\mu^2)$ for $i\neq j, (i,j\in\{1,2\})$. To
ensure the positivity of prices and demands, there have
$\beta_1\beta_2>\mu^2$ and $\alpha_i\beta_j>\alpha_j\mu$ for $i,j=1$
or $2$.

\subsection{Primary Users and Bertrand Game Model}

The spectra leased by primary users are either unused or transferred
from existing applications. The revenue of the $i^{th}$ primary user
is a product of the leased spectrum $q_i$ and the price $p_i$. In
practical cognitive radio networks, the primary users cannot always
satisfy the demands of the spectrum market. Hence, capacity
constraints should be taken into consideration when the primary
users set the prices. In this paper, we formulate the oligopoly
price competition as a Bertrand game. The \emph{players} are the
primary users, and the \emph{strategy} of a primary user is the
price of unit spectrum. Denote $\pi_i$ to be the \emph{payoff} (or
\emph{utility}) of the primary user $i$ that could be either the
revenue or the profit. Let $W_i$ be the spectrum size of the
$i^{th}$ primary user and $B_i$ be the traffic loads of primary
services. The spectrum efficiency, $r_i^{(p)}$, is defined as the
transmission rate per-unit spectrum for the $i^{th}$ primary
service. The available spectrum for sale at player $i$ is expressed
as $q_i^a = W_i-B_i/r_i^{(p)}$. We incorporate the capacity
constraints into the primary users' payoff via two approaches:

\begin{itemize}

\item \textbf{Type-I} \emph{Strict Capacity Constraints:} The
primary users aim to maximize their monetary revenues under the
constraints of capacities. The local optimization model of the
$i^{th}$ primary user is expressed as:
\begin{eqnarray}
\textrm{maximize}&&\pi_i(p_i,\mathbf{p}_{-i}) = p_i\cdot
\min\{f_i(\mathbf{p}), q_i^a\},\\
\textrm{subject to}&& p_i > 0, \;\;\; \forall i\in\mathcal{N},
\label{equation:sec2.7_type1model}
\end{eqnarray}
\noindent where $\mathbf{p}_{-i}$ is the price vector excluding
$p_i$.

\item \textbf{Type-II} \emph{QoS Penalty Functions:} An alternative
approach is to translate the capacity constraints as the barrier
penalty functions to the revenues. Assume that the traffic pattern
of a primary user is an i.d.d. poisson arrival process. The average
queuing delay of a packet can be approximated by
$\frac{B_i}{r_i^p(W_i-q_i)-B_i}$, which is adopted to reflect the
quality of primary services. Therefore, the utility maximization of
a primary user can be expressed as:
\begin{eqnarray}
\textrm{maximize} && \pi_i(p_i,\mathbf{p}_{-i}) =
p_i\min\{f_i(\mathbf{p}), q_i^a\} -\theta \log
\frac{B_i/r_i^{(p)}}{q_i^a-\min\{f_i(\mathbf{p}), q_i^a\}}
\label{equation:sec2.8_type2model}\\
\textrm{subject to} && p_i > 0, \;\;\; \forall i\in\mathcal{N},
\end{eqnarray}
\noindent where the positive variable $\theta$ is the weight of the
lognormal M/M/1 queuing delay.
\end{itemize}

Note that the capacity constraints in type-II have different
implication from that in type-I. There might have multiple
equilibrium points in the type-II model. Therefore, the infeasible
solutions are excluded if they are outside of the capacity bounds.
In the Bertrand game, the \emph{Nash Equilibrium} is a vector of
spectrum prices that no player can increase its payoff by changing
its price unilaterally.

\section{{\bf Noncooperative Game with Type-I Capacity Constraints}}
\label{section:type1}

In this section, we first present the static game and the
leader-follower game with \textbf{Type-I} capacity constraints, by
assuming the availability of full market information of primary
users. Furthermore, a dynamic game is formulated to characterize the
interactions of price competition when such information is not
available.

\subsection{Static Duopoly Game}
We commence the analysis by considering a duopoly spectrum market
with two primary users, and then extend to a more general scenario.
In the duopoly game, the \emph{NE} price of a player is obtained by
assuming that the other player also chooses the best strategy.
However, the ``\emph{best}'' strategies of primary users are
different in situations whether the spectrum capacities are
sufficient or not. Since there are two primary users, the
competitive pricing can be subdivided into four cases. Very
recently, authors have analyzed the static games of these four cases
and proved the existence of \emph{NE} in \cite{Mobihoc08:Zhang}.
Their analysis is based on graphical interpretation, which might be
difficult to extend to a more general oligopoly game. Inspired by
their work, we adopt a slightly different way to study the existence
of \emph{NE} in this section. In comparison with
\cite{Mobihoc08:Zhang}, the only difference is the simplicity of
analysis in this subsection. Later on, our method will be extended
to prove the existence of unique \emph{NE} in a spectrum game with
more than two primary users.

For the $i^{th}$ primary user, it decides the price $p_i$ so as to
maximize its revenue \cite{Mobihoc08:Zhang}
\begin{eqnarray}
\max \;\; p_i\cdot \min\{q_i^a, a_i-b_ip_i+cp_j\},\;\; \forall
i,j\in\{1,2\}. \label{equation:sec3.1.1}
\end{eqnarray}
\noindent We first investigate the Nash Equilibrium when the
available spectra are sufficient for both primary users (PUs). The
revenues of PU1 and PU2 can be written as:
\begin{eqnarray}
\pi_1(\mathbf{p}) = -b_1(p_1-\frac{a_1+cp_2}{2b_1})^2 +
\frac{(a_1+cp_2)^2}{4b_1},
\label{equation:sec3.1.2}\\
\pi_2(\mathbf{p}) = -b_2(p_2-\frac{a_2+cp_1}{2b_2})^2 +
\frac{(a_2+cp_1)^2}{4b_2}. \label{equation:sec3.1.3}
\end{eqnarray}
\noindent The best responses of PU1 and PU2 are:
\begin{eqnarray}
p_1^{*} = \frac{cp_2^{*}+a_1}{2b_1}, \;\;\; p_2^{*} =
\frac{cp_1^{*}+a_2}{2b_2}. \label{equation:sec3.1.4}
\end{eqnarray}
\noindent Thus, the duopoly prices at the unique \emph{NE} are:
\begin{eqnarray}
p_1^{*} = \frac{2a_1b_2+a_2c}{4b_1b_2-c^2}, \;\;\; p_2^{*} =
\frac{2a_2b_1+a_1c}{4b_1b_2-c^2}. \label{equation:sec3.1.5}
\end{eqnarray}
The spectrum demands at the unique \emph{NE} can be expressed by:
\begin{eqnarray}
q_1^{*} = \frac{2a_1b_1b_2+b_1a_2c}{4b_1b_2-c^2}, \;\;\; q_2^{*} =
\frac{2a_2b_1b_2+b_2a_1c}{4b_1b_2-c^2}. \label{equation:sec3.1.6}
\end{eqnarray}
The optimal response of the unconstrained game corresponds to the
\textbf{Case 1} that $q_1^a$ and $q_2^a$ satisfy $q_1^{a}>q_1^{*}$
and $q_2^{a}>q_2^{*}$. Three other cases are also considered when
the available spectra are not sufficient for the market demands.

\noindent \textbf{Case 2:} $q_1^{*}>q_1^{a}$ and $q_2^{a}$ is
sufficiently large. According to Eqn.(\ref{equation:sec3.1.1}), the
best revenue of PU1 is obtained at the point $p_1 =
\frac{a_1+cp_2}{2b_1}$ when the capacity is large enough
\begin{eqnarray}
q_1^a > \frac{a_1+cp_2}{2}. \label{equation:sec3.1.7}
\end{eqnarray}
\noindent Since we assume that $q_1^a$ is less than the best
spectrum demand, the revenue of PU1 is expressed as $\pi_1 = p_i
q_1^a$ if the following inequality holds:
\begin{eqnarray}
p_1\leq \frac{a_1-q_1^a+cp_2}{b_1}. \label{equation:sec3.1.8}
\end{eqnarray}
\noindent Here, we can easily find that this price bound is greater
than the best response when PU1 has a sufficiently large capacity.
We next analyze the selfish pricing behavior of PU1. Because PU1 can
lease at most $q_1^a$ units of spectrum, it is inclined to increase
$p_1$ for better monetary payoff until the spectrum demand is
exactly equal to the capacity. For PU2, the optimal response is
still characterized by Eqn.(\ref{equation:sec3.1.4}) so that it can
benefit from the increase of $p_1$. Hence, the prices at the Nash
Equilibrium can be solved by
\begin{eqnarray}
p_1^{\dag} = \frac{a_1-q_1^a+cp_2^{\dag}}{b_1}, \;\; \textrm{and}
\;\; p_2^{\dag} = \frac{a_2+cp_1^{\dag}}{2b_2}.
\end{eqnarray}
\noindent The results are give by
\begin{eqnarray}
p_1^{\dag}&=&\frac{2a_1b_2+a_2c-2b_2q_1^a}{2b_1b_2-c^2},
\label{equation:sec3.1.9}\\
p_2^{\dag}&=&\frac{a_2b_1+a_1c-cq_1^a}{2b_1b_2-c^2}.
\label{equation:sec3.1.10}
\end{eqnarray}
\noindent The spectrum demand of PU2 is:
\begin{eqnarray}
q_2^{\dag}=\frac{b_2(a_2b_1+a_1c-cq_1^a)}{2b_1b_2-c^2},
\label{equation:sec3.1.11}
\end{eqnarray}
\noindent while that of PU1 is exactly the capacity. Note that in
\emph{Case 2}, the capacity of PU2 must have $q_2^{\dag} \leq
q_2^a$.

\noindent \textbf{Case 3:} $q_2^{*}>q_2^{a}$ and $q_1^a$ is
sufficiently large. The leased spectrum of PU2 reaches $q_2^a$ so
that there exists
\begin{eqnarray}
p_2^{\ddag} = \frac{a_2-q_2^a+cp_1^{\ddag}}{b_2}.
\label{equation:sec3.1.12}
\end{eqnarray}
\noindent Following the same method in \emph{Case 2}, the \emph{NE}
prices of primary users are given by
\begin{eqnarray}
p_1^{\ddag} &=& \frac{a_1b_2+a_2c-cq_2^a}{2b_1b_2-c^2},
\label{equation:sec3.1.13} \\
p_2^{\ddag} &=& \frac{2a_2b_1+a_1c-2b_1q_2^a}{2b_1b_2-c^2}.
\label{equation:sec3.1.14}
\end{eqnarray}
\noindent The \emph{NE} demand of PU1 is expressed as:
\begin{eqnarray}
q_1^{\ddag} = \frac{b_1(a_1b_2+a_2c-cq_2^a)}{2b_1b_2-c^2}.
\label{equation:sec3.1.15}
\end{eqnarray}
\noindent Here, the capacity $q_1^a$ should be greater than
$q_1^{\ddag}$.

\noindent \textbf{Case 4:} There are two possible capacity sets in
this case, $q_1^{*}\geq q_1^{a}, q_2^{\dag}\geq q_2^{a}$ or
$q_1^{\ddag}\geq q_1^{a}, q_2^{*}\geq q_2^{a}$. When both PUs cannot
provide the best spectrum demands, they are disposed to increase the
prices until the spectrum demands equal to the capacities. Because
$q_1^a$ and $q_2^a$ are purchased by the secondary users, the prices
of per-unit spectrum can be obtained based on the following
equations:
\begin{eqnarray}
q_1^a = a_1 - b_1p_1^{\S} + cp_2^{\S}, \nonumber
\label{equation:sec3.1.16}\\
q_2^a = a_2 - b_2p_2^{\S} + cp_1^{\S}. \label{equation:sec3.1.17}
\end{eqnarray}
\noindent One can easily obtain the root of above equations:
\begin{eqnarray}
p_1^{\S} = \frac{a_1b_2+a_2c-b_2q_1^a-cq_2^a}{b_1b_2-c^2}, \nonumber
\label{equation:sec3.1.18}\\
p_2^{\S} = \frac{a_2b_1+a_1c-b_1q_2^a-cq_1^a}{b_1b_2-c^2}.
\label{equation:sec3.1.19}
\end{eqnarray}

\subsection{Static Oligopoly Game}
We study the existence of \emph{Nash Equilibrium} in a more
complicated spectrum market. Consider a set of primary users
$\mathcal{N}=\{1,2,\cdots N\}$ where each of them has a capacity
limit. Two key challenges hinder us from finding the existence of
\emph{NE}. First, we do not know which primary users have
insufficient capacities. Since the secondary users have preference
towards the primary users, a smaller capacity does not necessarily
mean the spectrum limitation compared with a larger one. Second, the
interaction of prices is still not well studied in the Bertrand
oligopoly market with capacity constraints.

To carry out our study, we recap some findings in the duopoly
spectrum market. A primary user is
\textbf{\emph{capacity-insufficient}} if the capacity is less than
the best demand with unlimited spectrum. The capacity-insufficient
primary user intends to increase the price until the capacity equals
to the market demand. However, we might not be able to find the
capacity-insufficient PUs once for all. When capacity-insufficient
PUs increases their prices of per-unit spectrum, secondary users may
go to other PUs, potentially leading to the lack of capacity in
those PUs. Therefore, we need to search several time recursively to
find the capacity-insufficient PUs. Inspired by the above findings,
we can obtain the \emph{NE} via the following steps. First, we
compute the best reactions of all PUs without considering the
capacity constraints. In this step, the $i^{th}$ PU decides the
price by
\begin{eqnarray}
p_i = \frac{a_i+\sum_{j\neq i}c_{ij}p_j}{2b_i}.
\label{equation:sec3.2.1_OligopolyPrice}
\end{eqnarray}
\noindent Denote $M_k$ to be the number of capacity-insufficient PUs
in the $k^{th}$ search. We can find $M_1$ primary users whose best
spectrum demands exceed the capacities in the first search. Let us
take the capacities of $M_1$ PUs into consideration. The
capacity-insufficient PUs have the incentive to increase their
prices so as to lower down the spectrum demands:
\begin{eqnarray}
p_i = \frac{a_i-q_i^a+\sum_{j\neq i}c_{ij}p_j}{b_i}.
\label{equation:sec3.2.2_OligopolyPriceCapaCons}
\end{eqnarray}
\noindent The remaining PUs increase the prices correspondingly. The
best reactions of primary users are solved through the equations in
Eqn.(\ref{equation:sec3.2.1_OligopolyPrice}) and
Eqn.(\ref{equation:sec3.2.2_OligopolyPriceCapaCons}). Define a new
matrix with the parameter $M_1$ as 
\begin{eqnarray}
\mathbf{Q}(M_1) = \left[ \begin{array}{ccccccc}
b_{1} & -c_{12} & \ldots & \ldots & \ldots & \ldots & -c_{1N} \\
-c_{21} & b_{2} & \ldots & \ldots & \ldots & \ldots & -c_{2N} \\
\vdots & \vdots & \ldots & \ldots & \ldots & \ldots & \vdots \\
-c_{M_11} & -c_{M_12}  & \ldots & b_{M_1} & \ldots & \ldots & -c_{M_1N}\\
-c_{M_1+1,1} & -c_{M_1+1,2}  & \ldots & \ldots & 2b_{M_1+1} & \ldots
& -c_{M_1+1,N}\\
\vdots & \vdots & \ldots & \ldots & \ldots & \ldots & \vdots\\
-c_{N1} & -c_{N2} & \ldots & \ldots & \ldots & \ldots & 2b_{N}
\end{array} \right]
\nonumber \label{equation:sec3.2.3_OligopolyPriceStage2}
\end{eqnarray}
\noindent and a vector
\begin{eqnarray}
\mathbf{a}(M_1) = [a_1-q_1^a \;\; a_2-q_2^a \;\; \cdots \;\;
a_{M_1}-q_{M_1}^{a} \;\; a_{M_1+1} \;\; \cdots \;\; a_N]^{T}.
\nonumber \label{equation:sec3.2.4_OligopolyPriceStage2_2}
\end{eqnarray}
\noindent Before identifying the capacity-insufficient PUs in the
next step, we need to know whether $\mathbf{Q}(M_1)$ is invertible
or not.
\begin{Lemma}
{\em The matrix $\mathbf{Q}(M_k)$ is positive definite if $\beta_i >
\mu > 0$ for all $i\in\mathcal{N}$ in the utility function.}
\label{Lemma:Invertible}
\end{Lemma}
\textbf{Proof:} Please refer to the Appendix. \done

\noindent Provided that $M_1$ primary users are
capacity-insufficient, the best responses can be solved via
\begin{eqnarray}
\mathbf{p} = \big[\mathbf{Q}(M_1)\big]^{-1}\cdot \mathbf{a}(M_1).
\label{equation:sec3.2.5_OligopolyPriceStage2_3}
\end{eqnarray}
\noindent For these $M_1$ PUs, their capacities and the best demands
have the following inequalities
\begin{eqnarray}
a_i+\sum_{j\neq i}c_{ij}p_j > 2q_i^a.
\label{equation:sec3.2.0_condforproof}
\end{eqnarray}
\noindent The spectrum demands of primary user can be obtained using
Eqn.(\ref{equation:sec2.4_demandfunceqn}). In the new solution
vector, we might observe that some additional primary users cannot
lease the best spectrum demands. As a result, they attempt to raise
the prices for better revenues. Assume that $M_2$ players have
limited capacities now, we replace the original $\mathbf{Q}(M_1)$
and $\mathbf{a}(M_1)$ by
\begin{eqnarray}
\mathbf{Q}(M_2) = \left[ \begin{array}{ccccccc}
b_{1} & -c_{12} & \ldots & \ldots & \ldots & \ldots & -c_{1N} \\
-c_{21} & b_{2} & \ldots & \ldots & \ldots & \ldots & -c_{2N} \\
\vdots & \vdots & \ldots & \ldots & \ldots & \ldots & \vdots \\
-c_{M_21} & -c_{M_22}  & \ldots & b_{M_2} & \ldots & \ldots & -c_{M_2N}\\
-c_{M_2+1,1} & -c_{M_2+1,2}  & \ldots & \ldots & 2b_{M_2+1} & \ldots
& -c_{M_2+1,N}\\
\vdots & \vdots & \ldots & \ldots & \ldots & \ldots & \vdots\\
-c_{N1} & -c_{N2} & \ldots & \ldots & \ldots & \ldots & 2b_{N}
\end{array} \right]
\nonumber \label{equation:sec3.2.6_OligopolyPriceStage3_1}
\end{eqnarray}
\noindent and
\begin{eqnarray}
\mathbf{a}(M_2) = [a_1-q_1^a \;\; a_2-q_2^a \;\; \cdots \;\;
a_{M_2}-q_{M_2}^{a} \;\; a_{M_2+1} \;\; \cdots \;\; a_N]^{T}.
\nonumber \label{equation:sec3.2.7_OligopolyPriceStage3_2}
\end{eqnarray}

Here, an important question is whether the iterative search method
can find more and more capacity-insufficient primary users? We must
show that the search method will not leads to a deadlock. Before
proving the nondecreasing property of search results, we introduce a
crucial definition first.
\begin{Definition}
{\em \textbf{Stieltjes matrix} \cite{Book03:Young}: A Stieltjes
matrix is a real symmetric positive definite matrix with nonpositive
off-diagonal entries. Every Stieltjes matrix is invertible to a
nonsingular symmetric matrix with nonnegative entries.}
\label{Def:Stieltjes}
\end{Definition}
\noindent According to Lemma \ref{Lemma:PositiveDefinite} and
\ref{Lemma:Invertible}, one can easily find that $\mathbf{Q}(M_k)$
is a Stieltjes matrix. In the following Lemma, we will show the
nondecreasing property of $M_k$.
\begin{Lemma}
{\em The set of capacity-insufficient primary users in the
$k-1^{th}$ step is a subset of that in the $k^{th}$ step.}
\label{Lemma:Nondecreasing}
\end{Lemma}
\textbf{Proof:} Please refer to the Appendix. \done

Using this method, we can find the primary users with capacity
shortage iteratively. The proposed method has low computational
complexity that requires at most $N$ searching steps. Next, we will
show that the oligopoly price vector computed above is a Nash
Equilibrium.

\begin{Theorem}
{\em The sets $\{M_k\}$ at subsequent steps of the search algorithm
form a nondecreasing sequence. The limit of which is the set $\{M\}$
such that the price vector computed by
\begin{eqnarray} \mathbf{p}^{*} = [\mathbf{Q}(M)]^{-1}\cdot \mathbf{a}(M).
\label{equation:sec3.2.8_OligopolyPriceStage4}
\end{eqnarray}
is a Nash Equilibrium of type-I oligopoly spectrum market.}
\label{Lemma:OligopolyNE}
\end{Theorem}

\noindent\textbf{Proof:} Assume that the primary users in the set
$\mathcal{M}=\{1,2,\cdots M\}$ have insufficient capacities in
respect to the best reactions. Their prices of per-unit spectrum are
are determined by
Eqn.(\ref{equation:sec3.2.2_OligopolyPriceCapaCons}). The remaining
primary users in the set $\mathcal{N}\setminus\mathcal{M}$ adjust
prices according to Eqn.(\ref{equation:sec3.2.1_OligopolyPrice}).
First, we show that any player $i\in\mathcal{N}\setminus\mathcal{M}$
has no incentive to adjust its price $p_i$. Define a new price
$p_i^{'} = p_i^{*}\pm\Delta$ where $\Delta$ is a positive deviation
from $p_i^{*}$ for $i\in\mathcal{N}\setminus\mathcal{M}$. The
difference of the revenues between $p_i^{*}$ and $p_i^{'}$ is
expressed as:
\begin{eqnarray}
&&\pi_i(p_i^{'},\mathbf{p}_{-i}^{*}) -
\pi_i(p_i^{*},\mathbf{p}_{-i}^{*}) \nonumber\\
&&= p_i^{'}(a_i-b_ip_i^{'}+\sum_{j\neq i}c_{ij}p_j^{*}) -
p_i^{*}(a_i-b_ip_i^{*}+\sum_{j\neq i}c_{ij}p_j^{*}) \nonumber\\
&&= (p_i^{*}\pm\Delta)(a_i-b_i(p_i^{*}\pm\Delta)+\sum_{j\neq
i}c_{ij}p_j^{*}) -
p_i^{*}(a_i-b_ip_i^{*}+\sum_{j\neq i}c_{ij}p_j^{*}) \nonumber\\
&&= -b_i\Delta^2 + (\pm\Delta)(a_i - 2b_ip_i^{*} + \sum_{j\neq
i}c_{ij}p_j^{*}) = -b_i \Delta^2 < 0.\nonumber
\end{eqnarray}
Thus, the $i^{th}$ primary user obtains smaller revenue if it
deviates from the NE price.

\noindent Next, we analyze the pricing strategies of the
capacity-insufficient PUs. In the iterative scheme, one principle is
that the best demand without capacity constraint is greater than
$q_i^a$ for any $i\in\mathcal{M}$ (i.e., in
Eqn.(\ref{equation:sec3.2.0_condforproof})). Consider the price
$p_i^{'} = p_i^{*}-\Delta$, the revenue of the $i^{th}$ PU is:
\begin{eqnarray}
\pi_i(p_i^{'},\mathbf{p}_{-i}^{*}) = q_i^a\times(p_i^{*}-\Delta) <
\pi_i(p_i^{*},\mathbf{p}_{-i}^{*}) \nonumber
\end{eqnarray}
\noindent because the spectrum demand $q_i^{'}$ exceeds the capacity
$q_i^a$. If the $i^{th}$ PU chooses a price $p_i^{'} =
p_i^{*}+\Delta$, the resulting revenue is:
\begin{eqnarray}
&&\pi_i(p_i^{'},\mathbf{p}_{-i}^{*}) =
(p_i^{*}+\Delta)(a_i-b_ip_i^{*}-b_i\Delta+\sum_{j\neq i}c_{ij}p_j)
\nonumber\\
&&= \pi_i(p_i^{*},\mathbf{p}_{-i}^{*}) +\Delta(a_i-2b_ip_i^{*}+
\sum_{j\neq i}c_{ij}p_j) - b_i\Delta^2 \nonumber \\
&&= \pi_i(p_i^{*},\mathbf{p}_{-i}^{*}) +\Delta(2q_i^a -(a_i +
\sum_{j\neq i}c_{ij}p_j)) - b_i\Delta^2 \nonumber \\
&&\leq\pi_i(p_i^{*},\mathbf{p}_{-i}^{*}) - b_i\Delta^2 <
\pi_i(p_i^{*},\mathbf{p}_{-i}^{*}). \nonumber
\end{eqnarray}
\noindent Since no player has the incentive to adjust its price, the
price vector computed by
Eqn.(\ref{equation:sec3.2.8_OligopolyPriceStage4}) is a Nash
Equilibrium of the oligopoly spectrum game. \done

Another key question is that whether the \emph{NE} found by our
method is unique or not. We have the following theorem for this
question:
\begin{Theorem}
{\em Consider a type-I oligopoly spectrum market in
Eqn.(\ref{equation:sec2.4_demandfunceqn}), there exists a unique
Nash Equilibrium.}\label{Lemma:UniqueNE}
\end{Theorem}
\textbf{Proof:} Please refer to the Appendix. \done

\subsection{Leader-Follower Duopoly Game}

Up to this point we have considered basically static spectrum games
in cognitive radio networks. However, the primary users may play
different roles in the price competition. We now extend the
noncooperative game to the leader-follower framework in which one
primary user moves first and then the other moves sequentially.
Consider a duopoly market, we show that the pricing strategies of
primary users rely on their leader/follower roles and capacities.
Similar to the static duopoly games, we also consider four cases in
the leader-follower games. (The notations
$p_i^{*},p_i^{\dag},p_i^{\ddag}$ and $p_i^{\S}$ are reused in this
subsection with superscripts $^l$ and $^f$ to stand for the roles of
PUs.)

\noindent \textbf{Case 1:} We consider the scenario that the PUs
have enough spectra for the secondary users. Let PU1 decide the
price first and the PU2 decide afterwards. They aim to maximize
their individual revenue. We use \emph{backward induction} to find
the subgame perfect \emph{NE} according to the best response of PU2,
$q_2^{f*}(q_1)$, for every possible value of $q_1$. Then, given that
PU1 knows PU2's best response, we obtain the best response of PU1.
Substitute $q_2$ by $q_2^{f*}(q_1)$ in the revenue function of PU1,
there has
\begin{eqnarray}
\pi_1(p_1,p_2^{f*}(p_1)) = p_1\cdot (a_1 - b_1p_1 +
\frac{c^2p_1+ca_2}{2b_2}) .
\end{eqnarray}
\noindent PU1 maximizes its revenue at the point:
\begin{eqnarray}
p_1^{l*} = \frac{2a_1b_2+ca_2}{2(2b_1b_2-c^2)}.
\end{eqnarray}
\noindent Then, the optimal price of PU2 is expressed as:
\begin{eqnarray}
p_2^{f*} = \frac{4a_2b_1b_2-a_2c^2+2a_1b_2c}{4b_2(2b_1b_2-c^2)}.
\end{eqnarray}
\noindent The spectrum demands of the primary users at the
leader-follower \emph{NE} are:
\begin{eqnarray}
q_1^{l*} &=& \frac{2a_1b_2+a_2c}{4b_2}, \\
q_2^{f*} &=&
\frac{4a_2b_1b_2^2-a_2b_2c^2+2a_1b_2^2c}{4b_2(2b_1b_2-c^2)}.
\end{eqnarray}
\noindent Note that the capacities $q_1^a$ and $q_2^a$ should be
greater than $q_1^{l*}$ and $q_2^{f*}$ respectively.

Similarly, using the backward induction method, we can also obtain
the price setting when PU2 is the leading service provider.
Comparing the leader-follower game with the static game, both the
leader and the follower achieve higher prices as well as revenues.

\noindent \textbf{Case 2:} $q_2^{a}$ is sufficiently large.
$q_1^{l*}>q_1^{a}$ if PU1 is the leader and $q_1^{f*}>q_1^{a}$ if
PU1 is the follower.

\begin{figure}[!htb]
\centering
\includegraphics[width = 3.5in]{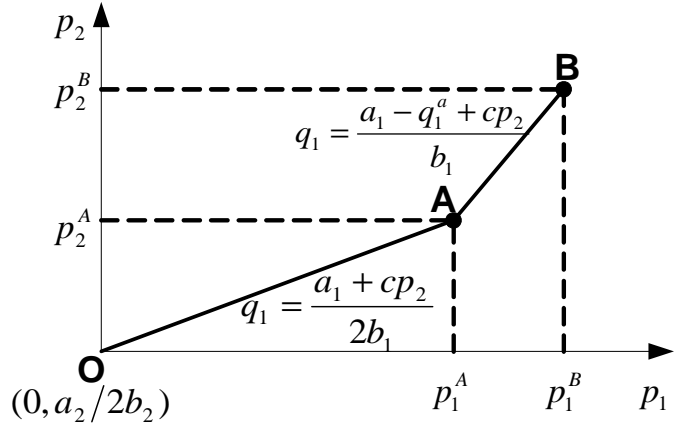}
\caption{\small{Interactions between $p_1$ and $p_2$}}
\label{figure:fig3.3.1_stackelberg}
\end{figure}

The leader-follower \emph{NE} depends on not only which primary user
decides the price first, but also whether the leader is
capacity-insufficient or not. As is mentioned in the static game,
the best response of PU1 is to set the price as $p_1 =
\max\{\frac{a_1+cp_2}{2b_1},\frac{a_1-q_1^a+cp_2}{b_1}\}$. To better
understand the leader-follower interaction, we make use of
Fig.\ref{figure:fig3.3.1_stackelberg} to illustrate the strategies
of primary users. Point $A$ represents the \emph{NE} in the static
game of \emph{Case 2}. Because the best response of PU1 is greater
than its capacity, it is inclined to increase $p_1$ for better
monetary payoff until $p_1^A$ is reached. In point $A$, the spectrum
demand of PU1 is exactly $q_1^a$ and the price is decided by
$p_1=\frac{a_1-q_1^a+cp_2}{b_1}$.

Let PU1 decide the price first and PU2 follow. As a leader, PU1
knows that the follower's best response is
$p_2^{f\dag}=\frac{a_2+cp_1}{2b_2}$. Then, in the first stage, the
revenue of PU1 is computed by:
\begin{eqnarray}
\pi_1 &=& \min\{q_1^a, a_1-b_1p_1+cp_2\}\cdot p_1 \noindent \\
&=& \min\{q_1^a, a_1-b_1p_1+\frac{c(a_2+cp_1)}{2b_2}\}\cdot p_1.
\end{eqnarray}
\noindent One can easily find that the best revenue is obtained at
the point $A$ if the leader has insufficient capacity. This is to
say, the \emph{NE} in the static game is also the \emph{NE} in the
leader-follower pricing when the capacity-insufficient PU acts as
the leader. Here, the capacity constraints must satisfy $q_1^a <
q_1^{l*}$ and $q_2^a > q_2^{f*}$ where $(q_1^{l*},q_2^{f*})$ is the
\emph{NE} in the unconstrained leader-follower market.

Next, we will show that the leader-follower spectrum game exhibits a
quite different strategy when PU2 is the leader. The PU2 has the
complete information of the PU1's best response in the second stage.
Because PU1 falls short of spectrum, it will set a higher price such
that the spectrum demand equals to the capacity: $p_1^{f\dag} =
\frac{a_1-q_1^a+cp_2}{b_1}$. Then, the revenue of PU2 is:
\begin{eqnarray}
\pi_2 = p_2\cdot (a_2-b_2p_2+\frac{c(a_1-q_1^a+cp_2)}{b_1}).
\end{eqnarray}
\noindent The best price of PU2 is thus given by:
\begin{eqnarray}
p_2^{l\dag}=\frac{a_1c+a_2b_1-cq_1^{a}}{2(b_1b_2-c^2)}.
\end{eqnarray}
\noindent Substitute $p_2$ by $p_2^{l\dag}$ in the best response of
$p_1^{f\dag}(p_2)$, we have the following expression of
$p_1^{f\dag}$:
\begin{eqnarray}
p_1^{f\dag} = \frac{2a_1b_1b_2-a_1c^2-2q_1^ab_1b_2+c^2q_1^a
+a_2b_1c}{2b_1(b_1b_2-c^2)}.
\end{eqnarray}
\noindent This leader-follower \emph{NE} is illustrated at point $B$
in Fig.\ref{figure:fig3.3.1_stackelberg}. Comparing the \emph{NE}
equilibria $A$ and $B$, we can see that the \emph{NE} prices depend
on the decision sequence of the primary users. When the primary user
with sufficient spectrum is the leader, both of them have higher
equilibrium prices. 
The purchased spectrum from the PU2 is given by
\begin{eqnarray}
q_2^{l\dag} =
\frac{a_2b_1^2b_2-a_2b_1c^2+a_1b_1b_2c-a_1c^3-q_1^ab_1b_2c+c^3q_1^a}{2b_1(b_1b_2-c^2)}.
\end{eqnarray}
\noindent Likewise, the capacity of PU2 must have $q_2^a \geq
q_2^{l\dag}$.

\noindent \textbf{Case 3:} $q_1^{a}$ is sufficiently large.
$q_2^{l*}>q_2^{a}$ if PU2 is the leader and $q_2^{f*}>q_2^{a}$ if
PU2 is the follower.

Following the method in \emph{Case 2}, we can obtain the Nash
Equilibria in the leader-follower games. We omit the solution
process and summarize the results as below.

\begin{itemize}
\item PU1 is the leader:
\begin{eqnarray}
p_1^{l\ddag} &=& \frac{a_2c+a_1b_2-cq_2^a}{2(b_1b_2-c^2)};\\
p_2^{f\ddag} &=&
\frac{2a_2b_1b_2-a_2c^2-2q_2^ab_1b_2+c^2q_2^a+a_1b_2c}{2b_2(b_1b_2-c^2)}.
\end{eqnarray}

\item PU2 is the leader:
\begin{eqnarray}
p_1^{f\ddag} &=& \frac{a_1b_2+a_2c-cq_2^a}{2b_1b_2-c^2}; \\
p_2^{l\ddag} &=& \frac{2a_2b_1+a_1c-2b_1q_2^a}{2b_1b_2-c^2}.
\end{eqnarray}
\end{itemize}

\noindent \textbf{Case 4:} The remaining capacity conditions exclude
the those in the other three cases. We adopt the backward induction
to find the leader-follower \emph{NE}s. Let PU1 be the leader and
PU2 be the follower. Give the price $p_1$, the best response of
player 2 is $p_2^{f\S} = \frac{a_2-q_2^a+cp_1}{b_2}$ if the capacity
$q_2^a$ is less than $\frac{a_2+cp_1}{2}$. Provided that PU1 knows
PU2's best response, we can obtain the best price of PU1,
$p_1^{l\S}(p_2^{f\S})$, so as to obtain the \emph{NE} for this game.
If PU1 is also capacity-limited, the price $p_1^{l\S}(p_2^{f\S})$ is
set to $\frac{a_1b_2+a_2c-b_2q_1^a-cq_2^a}{b_1b_2-c^2}$ and the best
price of PU2 is solved subsequently. The leader-follower game has
the same \emph{NE} as the static game. Furthermore, no matter which
primary user is the leader, the \emph{NE}s are the same when both of
them are capacity-insufficient.

\subsection{Dynamic Duopoly Game with Best Response Dynamics}

The best responses of the spectrum game are obtained under the
assumption that the primary users have a global knowledge of the
demand functions and the capacity constraints. However, the primary
users might only be able to observe the limited market information
in practice. A primary user may update its price in the next round
in response to the current prices of the opponents. In this
subsection, we investigate how primary users interact with each
other based on their individual best response functions.

In the duopoly Bertrand spectrum game, we assume that each primary
user merely knows its demand function, capacity and the price of the
opponent. Note that a player has no knowledge of the capacity of its
opponent. The spectrum price of the $i^{th}$ primary user at time
$t$ is denoted as $p_i(t)$, and that of the next slot is $p_i(t+1)$.
Here, the ``slot'' defines the length of time that primary users
adjust their prices. It can be one hour or one day, which is very
flexible. Hence, in each slot, player $i$ updates the price of
per-unit spectrum according to the best response function in the
static game:
\begin{eqnarray}
p_i(t+1) = \left\{\begin{array}{cc}
\frac{cp_j(t)+a_i}{2b_i} & \textrm{if $\frac{cp_j(t)+a_i}{2}\leq q_i^a$ for $i=1,2,$}\\
\frac{a_i-q_i^a+cp_j(t)}{b_i} & \textrm{if $\frac{cp_j(t)+a_i}{2}
\geq q_i^a$ for $i=1,2.$}
\end{array} \right.
\end{eqnarray}
We then study the stability by assuming the updating rules are
predetermined for the primary users.

\noindent \textbf{Case 1:} $q_1^{*}<q_1^{a}$ and $q_2^{*}<q_2^{a}$.
The update rule can be written in the matrix form:
\begin{eqnarray}
\mathbf{p}(t+1) = \left[ \begin{array}{cc}
0 & \frac{c}{2b_1} \\
\frac{c}{2b_2} & 0 \\
\end{array} \right]\cdot
\mathbf{p}(t) +  \left[ \begin{array}{c}
\frac{a_1}{2b_1} \\
\frac{a_2}{2b_2} \\
\end{array} \right]\;.
\label{equation:sec3.2.3_case1rule}
\end{eqnarray}
\noindent Denote $\lambda_1$ and $\lambda_2$ to be the eigenvalues
of the transfer matrix in Eqn.(\ref{equation:sec3.2.3_case1rule}).
One can easily find that $\lambda_1$ and $\lambda_2$ are within the
unit circle provided the feasibility constraint of the spectrum
game: $b_1b_2 > c^2$. According to the \emph{Routh-Hurvitz}
condition, the price adaptation scheme is stable.

\noindent \textbf{Case 2:} $q_1^{*}>q_1^{a}$ and $q_2^{*}<q_2^{a}$.
The updating rule in the matrix form is expressed as:
\begin{eqnarray}
\mathbf{p}(t+1) = \left[ \begin{array}{cc}
0 & \frac{c}{b_1} \\
\frac{c}{2b_2} & 0 \\
\end{array} \right]\cdot
\mathbf{p}(t) +  \left[ \begin{array}{c}
\frac{a_1-q_1^a}{b_1} \\
\frac{a_2}{2b_2} \\
\end{array} \right]\;.
\label{equation:sec3.4.3_case1rule}
\end{eqnarray}
\noindent Under the condition $b_1b_2 > c^2$, the eigenvalues of the
transfer matrix are also in the unit circle. Thus, the updating rule
is stable. The price updating schemes are stable in \textbf{Case 3}
and \textbf{Case 4}. We omit the analysis since they are similar to
\textbf{Case 1} and \textbf{Case 2}.

However, without the information of $q_i^{*}$, player $i$ might not
know how to select the price adaptation rule. We introduce a simple
algorithm named ``StrictBEST'' to update the prices without knowing
the demand function of the opponent. In each iteration, the $i^{th}$
primary user determines the price by
\begin{eqnarray}
p_i(t+1) =
\max\{\frac{cp_j(t)+a_i}{2b_i},\frac{a_i-q_i^a+cp_j(t)}{b_i}\}.
\label{equation:sec3.4.4_generalrule}
\end{eqnarray}
\noindent The above scheme has been used for the price adjustment of
the two-stage game in \cite{Mobihoc08:Zhang}. We present the
detailed proofs under different capacity constraints. However, the
above proof is incomplete because the primary users might switch the
price adjustment rules from time to time. This imposes great
difficulty to prove the convergence of the rule in
Eqn.(\ref{equation:sec3.4.4_generalrule}). Here, we present a
\emph{conjecture} on the \emph{StrictBEST} algorithm.
\begin{Conjecture}
{\em The \emph{StrictBEST} algorithm converges to the unique Nash
Equilibrium if the market parameters are positive as well as $b_1>c$
and $b_2 > c$.} \label{Conjecture:StrictBEST}
\end{Conjecture}

We propose a ``potential'' method to prove the above conjecture.
This method needs to consider many cases, which can not be exhausted
in this paper. An key observation is that the price of a primary
user is a function of its price two slot before. Thus, the
\emph{StrictBEST} algorithm might be able to converge if the prices
of primary users get closer and closer to the NE every 2 slots. As
we know, there are four types of capacity constraints (i.e.
\emph{Case 1$\sim$4} in this subsection). At time $t$, PU1 and PU2
have four different strategy profiles, and at time $t+1$, they also
have four types of adjustment strategies, resulting a total number
of 64 scenarios within two slots. Due to the complexity of this
``potential'' method, we do not prove the convergence property case
by case. Here, we abuse to denote the equilibrium price of $i^{th}$
PU to be $p_i^{*}$. The guidelines of the ``potential'' proof are
summarized as below,

\begin{itemize}

\item Prove that there exists only one equilibrium in
Eqn.(\ref{equation:sec3.4.4_generalrule}) when the spectrum
capacities $q_1^a$ and $q_2^a$ are given.

\item The distance between the price of user $i$ and its equilibrium
price is becoming smaller and smaller every two slots, that is,
$|p_i(t+2) - p_i^{*}| < |p_i(t) - p_i^{*}|$.

\end{itemize}

\subsection{Dynamic Duopoly Game with Bounded Rationality}

In a practical spectrum market, a primary user may not be able to
observe the profit gained by other primary services. Except the
adjustment rule based on best response function, a primary user can
also choose price for secondary users by learning the behaviors of
other players from the history. Bounded rationality mimics the human
behavior that players do not make perfectly rational decisions due
to the limited information and their conservativeness. The notion of
bounded rationality, also denoted as gradient dynamics, is employed
in dynamic Cournot oligopoly models
\cite{AOR99:Bischi,ADGA00:Bischi,NDSE05:Bischi}. Authors in
\cite{JSAC08:Niyato} adopts bounded rational strategy to adjust
spectrum price for the first time. The bounded rational rule is
equivalent to a distributed algorithm that gradually approaches the
equilibrium price. For instance, if a primary user is assumed to be
a bounded rational player, though ignorant of the actual demand, it
updates the price of per-unit spectrum based on a local estimate of
the marginal profit \cite{JSAC08:Niyato}. In this subsection, we
adopt the similar strategy to perform dynamic pricing in duopoly
spectrum market. The major difference lies in that we concentrate on
the interactions of duopoly primary users with capacity constraints.
When the capacity is large enough, the $i^{th}$ PU adjusts its price
according to the following rule:

\begin{eqnarray}
p_i(t+1) = p_i(t) + \gamma_i p_i(t)\cdot \frac{\partial
\pi_i(t)}{\partial p_i(t)},\;\;\; i=1,2,
\label{equation:sec3.5.1_boundedrationality}
\end{eqnarray}
\noindent where $\gamma_i$ is the learning stepsize. The learning
rate $p_i(t)\cdot \frac{\partial \pi_i(t)}{\partial p_i(t)}$
captures the conservativeness of players that may not totally
believe the observed market information. If player $i$ cannot
provide the optimal spectrum demand, it updates the price $p_i =
\frac{a_i-q_i^a+cp_j}{b_i}$ in each period.

In a self-mapping system, the most important issue is the stability
property. We first analyze local stability of the dynamic spectrum
sharing in \textbf{Case 1}. At the equilibrium point of price
adaptation, there has $\mathbf{p}(t+1)=\mathbf{p}(t) =
\mathbf{p^{*}}$. Since $p_i(t)$ is a self-mapping function, the
fixed points can be obtained by solving the following equations:
\begin{eqnarray}
\gamma_1 p_1(a_1-2b_1p_1+cp_2) = 0,
\label{equation:sec3.5.2_fixedpoint1}\\
\gamma_2 p_2(a_2-2b_2p_2+cp_1) = 0.
\label{equation:sec3.5.3_fixedpoint2}
\end{eqnarray}
\noindent In this duopoly spectrum game with bounded rationality,
the self-mapping dynamic system has four fixed points
$\mathbf{p}^{(1)}$, $\mathbf{p}^{(2)}$, $\mathbf{p}^{(3)}$ and
$\mathbf{p}^{(4)}$ as follows:

\begin{eqnarray}
&&\mathbf{p}^{(2)} = \big[\frac{a_1}{2b_1}, 0\big], \;\;\;
\mathbf{p}^{(3)} =
\big[0, \frac{a_2}{2b_2}\big], \nonumber \\
&&\mathbf{p}^{(4)} = \big[\frac{2a_1b_2+a_2c}{4b_1b_2-c^2},
\frac{2a_2b_1+a_1c}{4b_1b_2-c^2}\big], \nonumber
\end{eqnarray}
\noindent and $\mathbf{p}^{(1)}=[0,0]$. Clearly, one can see the
fixed point $\mathbf{p}^{(4)}$ without zero price is the Nash
Equilibrium.

Here, we apply the ``Routh-Hurvitz'' condition to analyze the
stability of these fixed points. At the equilibrium
$\mathbf{p}^{(1)}$, the Jacobian matrix of the self-mapping system
is expressed as:
\begin{eqnarray}
\mathbf{J}(\mathbf{p}^{(1)}) =\left[ \begin{array}{cc}
J_{11}(\mathbf{p}^{(1)}) & J_{12}(\mathbf{p}^{(1)}) \\
J_{21}(\mathbf{p}^{(1)}) & J_{22}(\mathbf{p}^{(1)}) \\
\end{array} \right]= \left[ \begin{array}{cc}
1+\gamma_1a_1 & 0 \\
0 & 1+\gamma_2a_2 \\
\end{array} \right]\;.
\label{equation:sec3.5.4_case1jacobp1}
\end{eqnarray}
\noindent The self-mapping system is stable only when the
eigenvalues of the Jacobian matrix are in the unit circle. Because
$\gamma_i$ and $a_i$ are nonnegative, the eigenvalues are greater
than 1. Hence, the fixed point $\mathbf{p}^{(1)}$ is unstable in the
self-mapping model. For the fixed point $\mathbf{p}^{(2)}$, the
Jacobian matrix is
\begin{eqnarray}
\mathbf{J}(\mathbf{p}^{(2)}) =\left[ \begin{array}{cc}
J_{11}(\mathbf{p}^{(2)}) & J_{12}(\mathbf{p}^{(2)}) \\
J_{21}(\mathbf{p}^{(2)}) & J_{22}(\mathbf{p}^{(2)}) \\
\end{array} \right] = \left[ \begin{array}{cc}
1-\gamma_1a_1 & \frac{\gamma_1a_1c}{2b_1} \\
0 & 1+\gamma_2(a_2+\frac{a_1c}{2b_1}) \\
\end{array} \right]\;
\label{equation:sec3.5.5_case1jacobp2}
\end{eqnarray}
\noindent which means that $\mathbf{p}^{(2)}$ is not a stable
equilibrium. Similarly, the fixed point $\mathbf{p}^{(3)}$ is not
stable either. For the fixed point $\mathbf{p}^{(4)}$, the Jacobian
matrix is
\begin{eqnarray}
\!\!\!\!\!\!\!\!\!\!\mathbf{J}(\mathbf{p}^{(4)}) \!=\!\left[
\begin{array}{cc}
J_{11}(\mathbf{p}^{(4)}) & J_{12}(\mathbf{p}^{(4)}) \\
J_{21}(\mathbf{p}^{(4)}) & J_{22}(\mathbf{p}^{(4)}) \\
\end{array} \right] \!=\! \left[ \begin{array}{cc}
1-\frac{2\gamma_1b_1(2a_1b_2+a_2c)}{4b_1b_2-c^2} & \frac{\gamma_1c(2a_1b_2+a_2c)}{4b_1b_2-c^2} \\
\frac{\gamma_2c(2a_2b_1+a_1c)}{4b_1b_2-c^2} & 1-\frac{2\gamma_2b_2(2a_2b_1+a_1c)}{4b_1b_2-c^2} \\
\end{array} \right].
\label{equation:sec3.5.6_case1jacobp4}
\end{eqnarray}
\noindent The characteristic function of the Jacobian matrix is
given by:
\begin{eqnarray}
\lambda^2 -
\lambda(J_{11}(\mathbf{p}^{(4)})+J_{22}(\mathbf{p}^{(4)})) -
J_{12}(\mathbf{p}^{(4)})\cdot J_{21}(\mathbf{p}^{(4)}) = 0.
\end{eqnarray}
\noindent The eigenvalues $\lambda_1$ and $\lambda_2$ are
\begin{eqnarray}
\lambda_{1,2} = \frac{(J_{11}+J_{22})\pm
\sqrt{(J_{11}+J_{22})^2-4J_{12}J_{21}}}{2} \;.
\label{equation:sec3.5.9_roots}
\end{eqnarray}
\noindent The fixed point $\mathbf{p}^{(4)}$ is stable only when
$|\lambda_1|$ and $|\lambda_2|$ are within the unit circle. Given
the spectrum market parameters and the learning rates
$\gamma_{1,2}$, one can easily check the stability of the Nash
Equilibrium.

Next, we analyze the dynamics of price adaptation with bounded
rationality in \textbf{Case 2}. Since PU1 cannot supply the best
spectrum demand of the secondary users, its price update follows the
rule $p_1=\frac{a_1-q_1^a+cp_2}{b_1}$. By letting
$\mathbf{p}(t+1)=\mathbf{p}(t)$, we solve the fixed points of the
self-mapping system: $\mathbf{p}^{(1)}=[\frac{a_1-q_1^a}{b_1},0]$
and $\mathbf{p}^{(2)}=[\frac{2b_2a_1+a_2c-2b_2q_1^a}
{2b_1b_2-c^2},\frac{a_1c+a_2b_1-q_1^ac}{2b_1b_2-c^2}]$. Likewise, we
apply the ``Routh-Hurvitz'' condition to the resulting Jacobian
matrices. For the fixed point $\mathbf{p}^{(1)}$, the Jacobian
matrix is given by
\begin{eqnarray}
\mathbf{J}(\mathbf{p}^{(1)}) = \left[ \begin{array}{cc}
0 & \frac{c}{b_1} \\
0 & 1+\gamma_2(a_2+\frac{c(a_1-q_1^a)}{b_1}) \\
\end{array} \right]\;.
\label{equation:sec3.5.12_case2jacobp1}
\end{eqnarray}
\noindent Note that $a_1$ is greater than the available spectrum
$q_1^a$. Thus, the fixed point $\mathbf{p}^{(1)}$ is unstable. In
the equilibrium point $\mathbf{p}^{(2)}$ of \emph{Case 2}, the
Jacobian matrix is written as
\begin{eqnarray}
\mathbf{J}(\mathbf{p}^{(2)}) = \left[ \begin{array}{cc}
0 & \frac{c}{b_1} \\
\frac{\gamma_2c(a_1c+a_2b_1-q_1^ac)}{2b_1b_2-c^2} &
1-\frac{2\gamma_2b_2(a_1c+a_2b_1-q_1^ac)}{2b_1b_2-c^2} \\
\end{array} \right]\;.
\label{equation:sec3.5.14_case2jacobp2}
\end{eqnarray}
\noindent Following Eqn.(\ref{equation:sec3.5.9_roots}), we can
obtain the eigenvalues $\lambda_1$ and $\lambda_2$, and validate the
stability of the self-mapping system. The stability analyses of
\textbf{Case 3} and \textbf{Case 4} are omitted because the
\textbf{Case 3} is very similar to the \textbf{Case 2} and the
\textbf{Case 4} adopts exactly the best response strategy.

In the dynamic spectrum game, a primary user has no information of
demand functions and capacities of its opponent. Thus, it is
necessary to compare the prices upon the situations whether the
capacity is sufficient or not. We present a distribute scheme,
namely ``StrictBR'', for the price update with strict capacity
constraints:
\begin{eqnarray}
\!\!\!\!\!\!\!\!\!\!p_i(t+1) =
\max\{\frac{a_i-q_i^a+cp_j(t)}{b_i},\;p_i(t) + \gamma_ip_i(t)
(a_i-2b_ip_i(t)+cp_j(t))\},
\end{eqnarray}
\noindent for $i,j = $ 1 or 2. When the price happens to be 0 in the
iteration, the primary users need choose a small positive price
randomly to leave the zero equilibrium point.

\section{{\bf Noncooperative Game with Type-II Capacity
Constraints}}\label{section:type2}

In this section, we analyze static and dynamic spectrum games with
\textbf{Type-II} capacity constraints. An iterative strategy is
proposed to set prices by using local market information.

\subsection{Static Duopoly Game}
\label{subsection:4.1}

We analyze the competitive pricing of duopoly primary users who aim
to maximize their utilities. The utility (or profit) is composed of
two parts, the \emph{revenue} and the \emph{delay-based cost}. In
the feasible region, the utility of the $i^{th}$ primary user is
expressed as
\begin{eqnarray}
\pi_i(p_i,\mathbf{p}_{-i}) = p_i (a_i-b_ip_i+cp_j) -\theta \log
\frac{B_i/r_i^{(p)}}{q_i^a-a_i+b_ip_i-cp_j}, \;\;\; \forall i,j\in
\{1,2\}. \nonumber
\end{eqnarray}
\noindent To find the best response of the primary users, we
differentiate the utility $\pi_i$ with respect to $p_i$ and let the
derivative be 0, there have
\begin{eqnarray}
\frac{\partial \pi_1}{p_1} = a_1 - 2b_1p_1 +cp_2
+\frac{\theta b_1}{q_1^a-a_1+b_1p_1-cp_2} = 0,
\label{equation:sec4.1.1_type2diff}\\
\frac{\partial \pi_2}{p_2} = a_2 - 2b_2p_2 + cp_1 +\frac{\theta
b_2}{q_2^a-a_2+b_2p_2-cp_1} = 0. \label{equation:sec4.1.2_type2diff}
\end{eqnarray}
\noindent To reduce the complexity of expression in
Eqn.(\ref{equation:sec4.1.1_type2diff}) and
(\ref{equation:sec4.1.2_type2diff}), we represent the prices $p_i$
using the spectrum demands $q_i$ in
Eqn.(\ref{equation:sec2.2_pricefunc}) for $i=1,2$. Thus, the above
equations are transformed into follows:
\begin{eqnarray}
q_1 - b_1(\alpha_1-\beta_1q_1-\mu q_2) + \frac{\theta
b_1}{q_1^a-q_1} = 0,
\label{equation:sec4.1.3_type2diff}\\
q_2 - b_2(\alpha_2-\beta_2q_2-\mu q_1) + \frac{\theta
b_2}{q_2^a-q_2} = 0, \label{equation:sec4.1.4_type2diff}
\end{eqnarray}
\noindent where $q_i$ is within the feasible region
$\mathbb{S}=\{q_i|0\leq q_i\leq q_i^a, i=1,2\}$. In the above
nonlinear equations, the explicit forms of $q_1$ and $q_2$ cannot be
solved directly. Submit $q_1$ in
Eqn.(\ref{equation:sec4.1.3_type2diff}) to
Eqn.(\ref{equation:sec4.1.4_type2diff}), we have
\begin{eqnarray}
&&\frac{b_1\alpha_1-(1+\beta_1b_1)q_1-\frac{\theta
b_1}{q_1^a-q_1}}{\mu b_1}\cdot (1+\beta_2b_2) + \mu b_2q_1 \nonumber \\
&&= b_2\alpha_2 -\frac{\theta b_2}{q_2^a -
\frac{b_1\alpha_1-(1+\beta_1 b_1)q_1 -\frac{\theta b_1}{q_1^a -
q_1}}{\mu b_1}}. \label{equation:sec4.1.5_q1q2equation}
\end{eqnarray}
\noindent We next show that there is only one solution in the range
$(0, q_1^a)$. The left-hand expression can be further rewritten as:
\begin{eqnarray}
&&\frac{b_1\alpha_1-q_1-\frac{\theta b_1}{q_1^a-q_1}}{\mu b_1}\cdot
(1+\beta_2b_2) -
\frac{1}{\mu}(\beta_1+\beta_1\beta_2b_2-\mu^2b_2).\nonumber
\end{eqnarray}
\noindent One can see that the left-hand expression is a strictly
decreasing function of $q_1$ in the feasible region. The right-hand
expression of Eqn.(\ref{equation:sec4.1.5_q1q2equation}) is a
strictly increasing function of $q_1$ in the range $(0, q_1^a)$.
When $q_1$ approaches $q_1^a$, the right-hand expression is
approximated by $b_2\alpha_2$, while the left-hand expression is
negatively infinite. Hence, there exist a unique feasible solution
in Eqn.(\ref{equation:sec4.1.5_q1q2equation}) only if the left-hand
is greater than the right-hand at the point $q_1 = 0$. We can find
the range of $\theta$ to guarantee the unique feasible solution for
$q_1$. Here, we only show that there exists a unique $q_1\in
(0,q_1^a)$ in this duopoly market when $\theta$ is small. In the
point $q_1 = 0$, the difference between the left-hand and the
right-hand is approximated by
\begin{eqnarray}
\frac{a_1}{\mu} + \frac{\alpha_1\beta_2b_2 - \alpha_2 b_2 \mu}{\mu}
> 0\nonumber
\end{eqnarray}
\noindent since $\alpha_1\beta_2 > \alpha_2\mu$ holds in the duopoly
model. The above analysis also implies that there exists a unique
Nash Equilibrium when $\theta$ is sufficiently small.

\subsection{Static Oligopoly Game}
\label{subsection:4.2}

In this subsection, we extend the above analysis to a more general
oligopoly spectrum game with \emph{type-II} constraints. By
introducing a novel variable transformation method, we analytically
show the existence of unique \emph{NE}. Given the utility of the
$i^{th}$ primary user, we take the first-order derivative over
$p_i$:
\begin{eqnarray}
\frac{\partial \pi_i}{p_i} = a_i - 2b_ip_i +c\sum_{j\neq i}p_j
+\frac{\theta b_i}{q_i^a-a_i+b_ip_i-c\sum_{j\neq i}p_j} = 0.
\label{equation:sec4.2.1_type2diff}
\end{eqnarray}
\noindent Unlike the duopoly spectrum game, we cannot directly find
the equations to solve $p_i$ for more than two primary users. To
simplify the analysis, we substitute the variables $p_i$ by $q_i$
according to Eqn.(\ref{equation:sec2.2_pricefunc}) for
$i\in\mathcal{N}$. Formally, there exists
\begin{eqnarray}
q_i - b_i(\alpha_i-\beta_iq_i -\mu\sum_{j\neq i}p_j) +\frac{\theta
b_i}{q_i^a-q_i} = 0, \;\; \forall i\in\mathcal{N}.
\label{equation:sec4.2.2_type2diff}
\end{eqnarray}
\noindent We denote a new variable $Z = \sum_{i\in\mathcal{N}}q_i$
to be the total spectrum provision in the oligopoly market. Thus,
the above equations can be transformed into:
\begin{eqnarray}
(1+\beta_ib_i-\mu b_i)q_i + \mu b_i Z -\alpha_ib_i +
\frac{\theta_ib_i}{q_i^a-q_i} = 0,\;\; \forall i\in\mathcal{N}.
\label{equation:sec4.2.3_type2diff}
\end{eqnarray}
\noindent Assume that $Z$ is a constant, the original coupled
equations are converted into a set of separated quadratic
formations:
\begin{eqnarray}
&&(1+\beta_ib_i-\mu b_i)q_i^2 - ((1+\beta_ib_i-\mu b_i)q_i^a +
\alpha_ib_i-\mu b_i Z)q_i \nonumber \\
&&-(\mu b_i Z q_i^a - \alpha_i b_i q_i^a + \theta b_i) = 0,\;\;
\forall i\in\mathcal{N}. \label{equation:sec4.2.4_type2diff}
\end{eqnarray}
\noindent For the $i^{th}$ PU, $q_i$ has two roots:
\begin{eqnarray}
&&\!\!\!\!\!\!\!\!\!\!q_i^{(1)} = \frac{((1+\beta_ib_i-\mu b_i)q_i^a
+ \alpha_ib_i-\mu b_i Z)}{2(1+\beta_ib_i-\mu
b_i)}\nonumber\\
&&\!\!\!\!\!\!\!\!\!\!+ \frac{\sqrt{((1+\beta_ib_i-\mu b_i)q_i^a +
\alpha_ib_i-\mu b_i Z)^2+4(1+\beta_ib_i-\mu b_i)(\mu b_i Z q_i^a -
\alpha_i b_i q_i^a +
\theta b_i)}}{2(1+\beta_ib_i-\mu b_i)} \nonumber\\
&&\!\!\!\!\!\!\!\!\!\!=\frac{((1+\beta_ib_i-\mu b_i)q_i^a +
\alpha_ib_i-\mu b_i Z)}{2(1+\beta_ib_i-\mu
b_i)} \nonumber\\
&&\!\!\!\!\!\!\!\!\!\!+ \frac{\sqrt{(-(1+\beta_ib_i-\mu b_i)q_i^a +
\alpha_ib_i-\mu b_i
Z)^2+4(1+\beta_ib_i-\mu b_i) \theta b_i}}{2(1+\beta_ib_i-\mu b_i)}\\
&&\!\!\!\!\!\!\!\!\!\!q_i^{(2)} = \frac{((1+\beta_ib_i-\mu b_i)q_i^a
+ \alpha_ib_i-\mu b_i Z)}{2(1+\beta_ib_i-\mu
b_i)} \nonumber\\
&&\!\!\!\!\!\!\!\!\!\!- \frac{\sqrt{(-(1+\beta_ib_i-\mu b_i)q_i^a +
\alpha_ib_i-\mu b_i Z)^2+4(1+\beta_ib_i-\mu b_i) \theta
b_i}}{2(1+\beta_ib_i-\mu b_i)}. \label{equation:sec4.2.10}
\end{eqnarray}
\noindent We need to validate if the roots are within the feasible
ranges. Let $q_i^{(1)}$ and $q_i^{(2)}$ minus $q_i^a$, we have
\begin{eqnarray}
&&q_i^{(1)} - q_i^a = \frac{(-(1+\beta_ib_i-\mu b_i)q_i^a +
\alpha_ib_i-\mu b_i Z)}{2(1+\beta_ib_i-\mu b_i)} \nonumber \\
&&+ \frac{\sqrt{(-(1+\beta_ib_i-\mu b_i)q_i^a + \alpha_ib_i-\mu b_i
Z)^2+4(1+\beta_ib_i-\mu b_i) \theta b_i}}{2(1+\beta_ib_i-\mu b_i)}>
0,
\end{eqnarray}
\noindent and
\begin{eqnarray}
&&q_i^{(2)} - q_i^a = \frac{(-(1+\beta_ib_i-\mu b_i)q_i^a +
\alpha_ib_i-\mu b_i Z)}{2(1+\beta_ib_i-\mu b_i)} \nonumber \\
&&- \frac{\sqrt{(-(1+\beta_ib_i-\mu b_i)q_i^a + \alpha_ib_i-\mu b_i
Z)^2+4(1+\beta_ib_i-\mu b_i) \theta b_i}}{2(1+\beta_ib_i-\mu b_i)}<
0,
\end{eqnarray}
\noindent because $\beta_i$ is greater than $\mu$. When $q_i$ is
chosen to be $q_i^{(2)}$ for each primary user $i$, it is
represented by a function of $Z$. Define a set of functions
$h_i(Z)=q_i^{(2)}$ for $i\in \mathcal{N}$, we will show that
$h_i(Z)$ is a decreasing function. Differentiate $h_i(Z)$ over $Z$,
we have
\begin{eqnarray}
\!\!\!\!\!\!\!\!\!\!&&\frac{\partial h_i(Z)}{\partial Z} = \nonumber \\
\!\!\!\!\!\!\!\!\!\!&&\frac{-\mu b_i\sqrt{(-(1+\beta_ib_i-\mu
b_i)q_i^a + \alpha_ib_i-\mu b_i Z)^2+4(1+\beta_ib_i-\mu b_i) \theta
b_i}}{2(1+\beta_ib_i-\mu b_i)(\sqrt{(-(1+\beta_ib_i-\mu b_i)q_i^a +
\alpha_ib_i-\mu b_i
Z)^2+4(1+\beta_ib_i-\mu b_i) \theta b_i})} \nonumber \\
\!\!\!\!\!\!\!\!\!\!&&\frac{\mu b_i (-(1+\beta_ib_i-\mu b_i)q_i^a +
\alpha_ib_i-\mu b_i Z)}{2(1+\beta_ib_i-\mu
b_i)(\sqrt{(-(1+\beta_ib_i-\mu b_i)q_i^a + \alpha_ib_i-\mu b_i
Z)^2+4(1+\beta_ib_i-\mu b_i) \theta b_i})}
\nonumber \\
\!\!\!\!\!\!\!\!\!\!&&< 0.
\end{eqnarray}
\noindent Sum up all the spectrum demands, we obtain the following
self-mapping equation:
\begin{eqnarray}
Z = \sum_{i=1}^{N}h_i(Z). \label{equation:sec4.2.14_fixedpointEqn}
\end{eqnarray}
\noindent The left-hand of the above one-dimensional function is
strictly increasing, while the right-hand is a bounded and
decreasing function. To guarantee the existence of unique solution,
the right-hand should be greater than the left-hand when $Z$ is 0:
\begin{eqnarray}
&&\!\!\!\!\!\!\!\!\!\!\sum_{i=1}^{N}\frac{((1+\beta_ib_i-\mu
b_i)q_i^a +
\alpha_ib_i)}{2(1+\beta_ib_i-\mu b_i)} \nonumber \\
&& - \sum_{i=1}^{N}\frac{\sqrt{(-(1+\beta_ib_i-\mu b_i)q_i^a +
\alpha_ib_i)^2+4(1+\beta_ib_i-\mu b_i) \theta
b_i}}{2(1+\beta_ib_i-\mu b_i)}> 0. \label{equation:sec4.2.15}
\end{eqnarray}
\noindent The above inequality has a much simpler necessary
condition. By separating the inequality (\ref{equation:sec4.2.15})
into $N$ smaller inequalities for primary users, we obtain the
necessary conditions $\alpha_i q_i^a \geq \theta$ for $i\in
\mathcal{N}$. The solution $Z^{*}$ can be solved numerically via
binary search or golden search methods. Subsequently, the individual
spectrum demands can be obtained through
Eqn.(\ref{equation:sec4.2.10}). Since the demand $q_i$ is
nonnegative, $Z^{*}$ and the market parameters must have $\mu b_i Z
q_i^a - \alpha_i b_i q_i^a + \theta b_i \leq 0$. Formally, we have
the following theorem on the existence of unique \emph{NE}:
\begin{Theorem}
{\em Consider a type-II oligopoly spectrum market in
Eqn.(\ref{equation:sec2.8_type2model}) with $N$ primary users. There
exists a unique Nash Equilibrium if the following conditions hold:
\begin{itemize}
\item $\beta_i > \mu$ for all $i\in\mathcal{N}$;

\item $\alpha_i,\beta_i,\mu,a_i,b_i > 0$ for all $i\in\mathcal{N}$,
and $c_{i,j} > 0$ for $i,j\in\mathcal{N}, i\neq j$;

\item $\alpha_i q_i^a \geq \theta$ for $i\in
\mathcal{N}$;

\item Given the unique solution $Z^{*}$ to
Eqn.(\ref{equation:sec4.2.14_fixedpointEqn}), there has $\mu b_i Z
q_i^a - \alpha_i b_i q_i^a + \theta b_i \leq 0$ for
$i\in\mathcal{N}$.

\end{itemize}}
\end{Theorem}

\subsection{Dynamic Duopoly Game with Best Response Dynamics}

In the dynamic spectrum game, the revenue and the QoS of a primary
user are not available to its opponent. Hence, the decision of
prices is made based on the local utility function and the observed
prices. We adopt the best response scheme to adjust the prices of
per-unit spectrum. Each player assumes that it is the only service
provider in the spectrum market. According to
Eqn.(\ref{equation:sec4.1.1_type2diff}) and
(\ref{equation:sec4.1.2_type2diff}), the price $p_i$ is a function
of $p_j \; (\forall i\neq j)$ in each time period
\begin{eqnarray}
&&\!\!\!\!\!\!\!\!\!\! 2b_1^2p_1^2 +
\big(2b_1(q_1^a-a_1-cp_2)-b_1(a_1+cp_2)\big)p_1
\nonumber\\
&&
\;\;\;\;\;\;\;\;\;\;\;\;\;\;\;\;\;\;\;\;\;\;\;\;\;-(a_1+cp_2)(q_1^a-a_1-cp_2)
-b_1\theta = 0,
\label{equation:sec4.3.1}\\
&&\!\!\!\!\!\!\!\!\!\! 2b_2^2p_2^2 +
\big(2b_2(q_2^a-a_2-cp_1)-b_2(a_2+cp_1)\big)p_2
\nonumber\\
&&
\;\;\;\;\;\;\;\;\;\;\;\;\;\;\;\;\;\;\;\;\;\;\;\;\;-(a_2+cp_1)(q_2^a-a_2-cp_1)
-b_2\theta = 0. \label{equation:sec4.3.2}
\end{eqnarray}
\noindent The prices $p_1$ and $p_2$ can be easily solved by
treating the opponent's price as a constant. Denote $p_1^{(1)}(p_2)$
and $p_1^{(2)}(p_2)$ to be the roots of
Eqn.(\ref{equation:sec4.3.1}). Let $p_2^{(1)}(p_1)$ and
$p_2^{(2)}(p_1)$ be the roots of Eqn.(\ref{equation:sec4.3.2}). A
subsequent problem is how to select the ``appropriate'' prices for
both players. According to the
Eqn.(\ref{equation:sec2.6_demandprice}) and
(\ref{equation:sec2.7_demandprice}), player $i$ computes the
spectrum demands $q_i(p_i^{(1)}(p_j))$ and $q_i(p_i^{(2)}(p_j))$ for
$i\neq j$. If one of the demands is in the range $(0,q_i^a)$, the
corresponding price is selected to lease the spectrum. The preceding
analysis also shows that under certain market condition, there is
only one feasible price for each primary user. Both of the demands
might also be outside/inside of the range $(0,q_i^a)$ for some
special reasons such as random noise etc. Under this situation, the
price is chosen to receive better utility. We propose a distributed
algorithm named ``QoSBEST'' to determine the prices, which can also
be extended to a more general oligopoly market. The ``QoSBEST''
algorithm is specified in Fig.\ref{figure:QoSBEST}.
\begin{figure}[htb]
\center
\begin{small}
\begin{tabbing}
\hskip 0.12in \=xx\=x\=x\=x\=x\=x\=x\=x\kill\\
\rule{4.5in}{0.25mm}\\
1: {\em Calculate the prices of per-unit spectrum in Eqn.(\ref{equation:sec4.3.1}) and (\ref{equation:sec4.3.2})};\\
2: {\em $q_i^{(k)}(t+1) = a_i-b_ip_i^{(k)}(t+1) +cp_j(t),\;\;$ for
$i\neq j$ and
$k=1,2$}; \\
\> //predict the current demands\\
3: {\bf If} {\em $q_i^{(1)} < q_i^a$ and $q_i^{(2)}\geq q_i^a$}\\
4: \> \> {\em $p_i(t) = p_i^{(1)}$};\\
5: {\bf elseif} {\em $q_i^{(1)} \geq q_i^a$ and $q_i^{(2)} < q_i^a$}\\
6: \> \> {\em $p_i(t) = p_i^{(2)}$};\\
7: {\bf end}\\
\rule{4.5in}{0.25mm}
\end{tabbing}
\end{small}
\vspace{-0.15in} \caption{\sf QoSBEST Algorithm}
\label{figure:QoSBEST}
\end{figure}

An important question is that whether the \emph{QoSBEST} algorithm
converges to the Nash Equilibrium. We first give a lemma that will
be used in the proof and then present the complete proof of
convergence.

\begin{Lemma}
{\em For any real positive constants $x$, $y$ and $z$, there have
\begin{eqnarray}
-(x+y) \leq \sqrt{x^2+z^2} - \sqrt{y^2+z^2} \leq x+y.
\label{equation:sec4.3.5}
\end{eqnarray}
and
\begin{eqnarray}
\sqrt{x^2+z^2} - \sqrt{y^2+z^2} \leq x - y, \;\;\;\;\; \textrm{if}
\;\; x\geq y. \label{equation:sec4.3.6}
\end{eqnarray}}
\label{Lemma:Inequality}
\end{Lemma}
\textbf{Proof:} Omitted due to its simplicity. \done

\begin{Theorem}
{\em The \emph{QoSBEST} algorithm converges to the unique NE if the
market parameters are positive as well as $b_1>c > 0$ and $b_2 > c
> 0$.} \label{Theorem:QoSBEST_Converge}
\end{Theorem}
\textbf{Proof:} Please refer to the Appendix. \done

With the \emph{type-II} constraints, there are multiple fixed points
in the bounded rationality model. The marginal profit based
iterative scheme may possibly converge to a fixed point outside of
the feasible region. Hence, we avoid to consider the dynamic game
with bounded rationality in this section.

\section{{\bf Performance Evaluation}}
\label{section:evaluation}

We present the numerical results to evaluate the price competition
and the performance of adjustment strategies in the duopoly spectrum
market. Distinguished from some previous work, several important
dynamic behaviors are investigated such as \emph{bifurcation
diagrams, strange attractors and Lyapunov exponents}.

\subsection{Static and Leader-Follower Games with Type-I Constraints}

\begin{figure}[htb]
    \centering
    \includegraphics[width=3.5in]{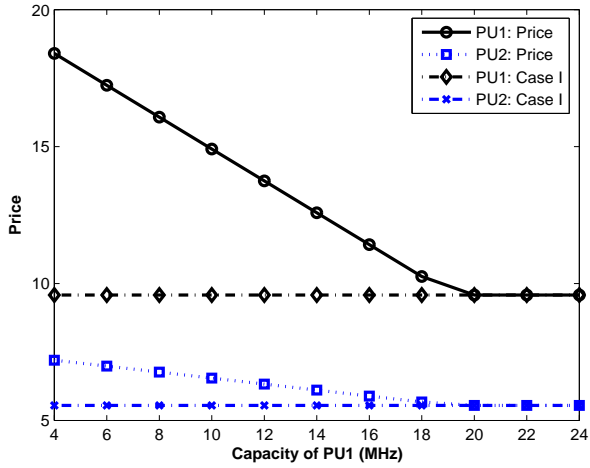}
    \caption{The change of prices when $q_1^a$ increases from 4MHz
    to 24MHz and $q_2^a$ is sufficiently large}
    \label{figure:pricevspu1c}
\end{figure}

\begin{figure}[!htb]
    \centering
    \includegraphics[width=3.5in]{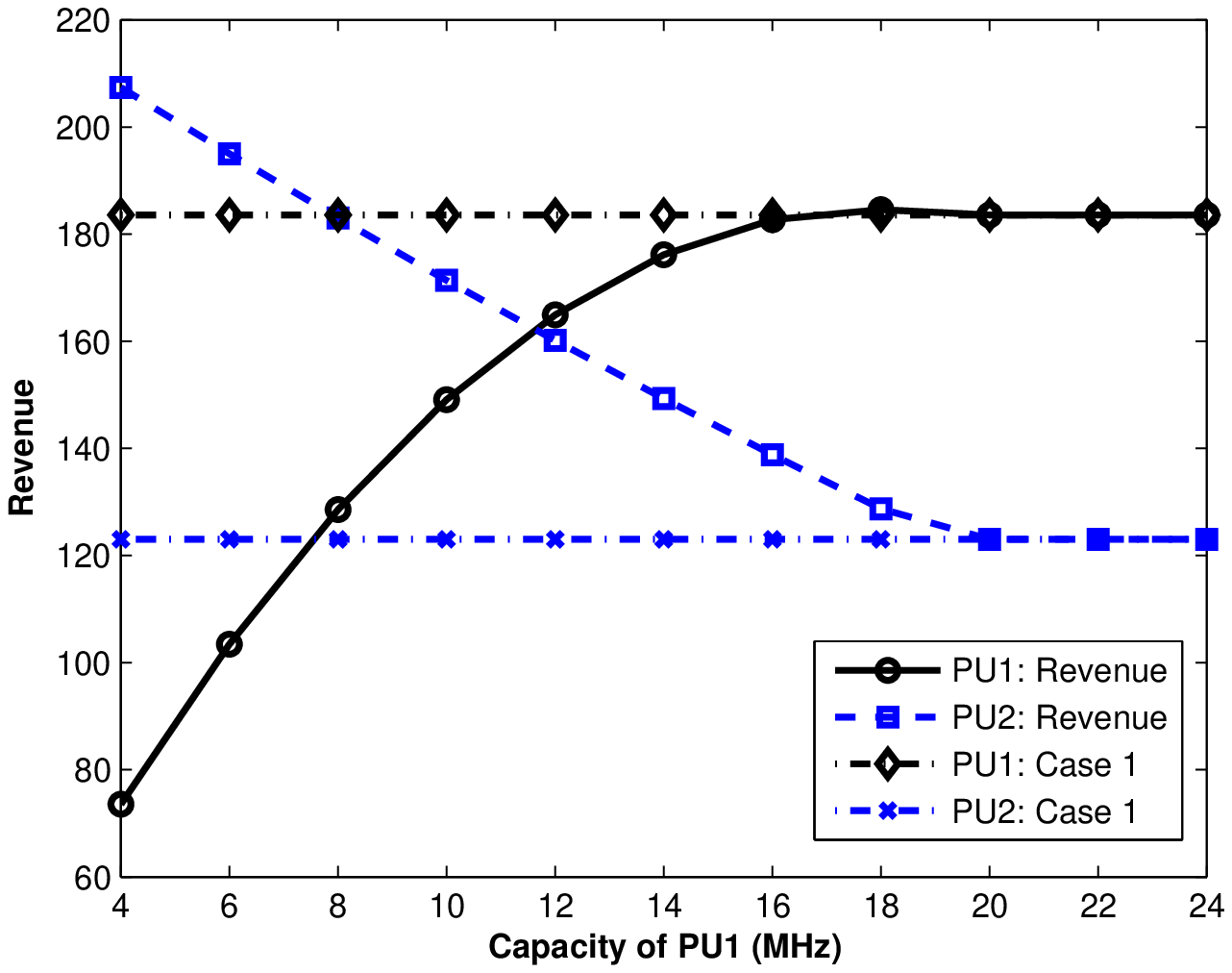}
    \caption{The change of revenues when $q_1^a$ increases from 4MHz
    to 24MHz and $q_2^a$ is sufficiently large}
    \label{figure:revenuevspu1c}
\end{figure}

\begin{figure}[!htb]
    \centering
    \includegraphics[width=3.5in]{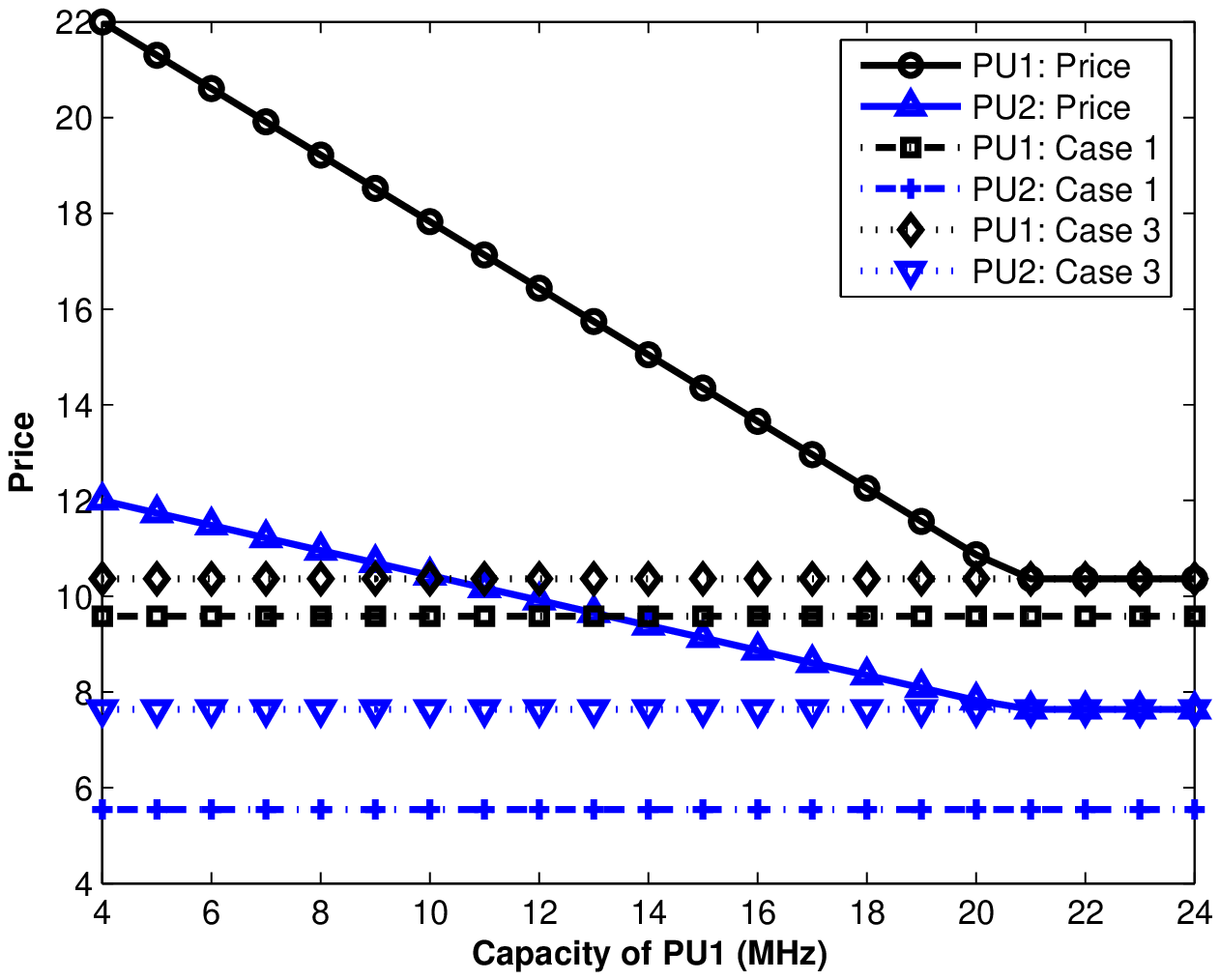}
    \caption{The change of prices when $q_1^a$ increases from 4MHz
    to 24MHz and $q_2^a$ is 15MHz}
    \label{figure:pricevsp1_p2limited}
\end{figure}

\begin{figure}[!htb]
    \centering
    \includegraphics[width=3.5in]{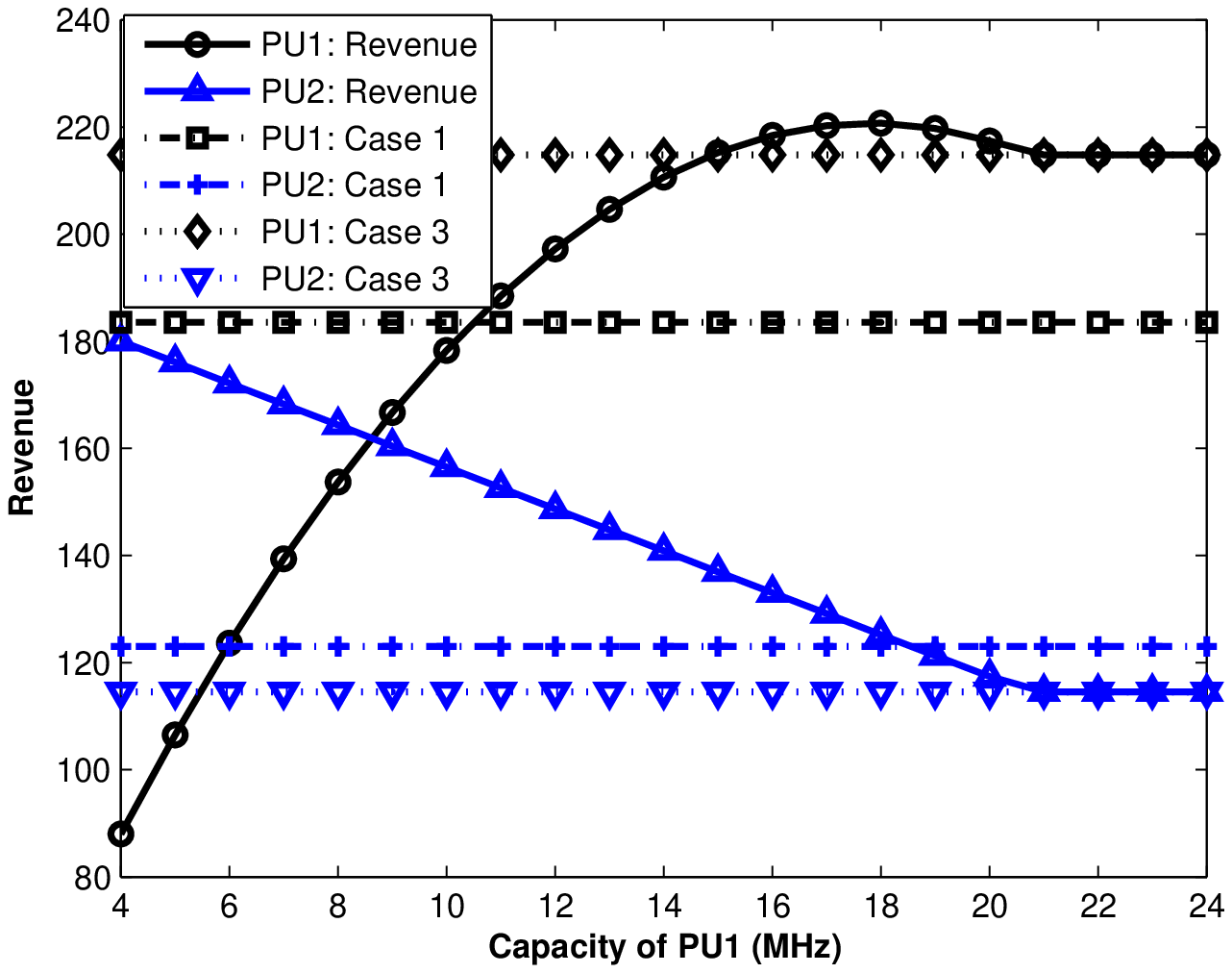}
    \caption{The change of revenues when $q_1^a$ increases from 4MHz
    to 24MHz and $q_2^a$ is 15MHz}
    \label{figure:revenuevsp1_p2limited}
\end{figure}

We first study the \emph{NE}s of static and leader-follower duopoly
games under different settings. To perform the numerical analysis,
we configure the parameters for the Bertrand demand functions:
\begin{eqnarray}
q_1 &=& 30MHz - 2p_1 + 1.5p_2; \nonumber \\
q_2 &=& 30MHz-4p_2+1.5p_1. \nonumber
\end{eqnarray}
\noindent where the unit of $p_i$ is dollar/MHz and that of $q_i$ is
MHz for $i=1,2$. In this cognitive radio environment, the primary
users have capacity constraints when leasing the unused spectra.
According to the best responses in the \emph{Type-I} model, one can
easily compute the optimal prices and spectrum demands without
capacity constraints: $p_1^{*}=9.58, p_2^{*}=5.55, q_1^{*}= 19.16$
and $q_2^{*}=22.18$. In the first experiment, we assume that the
capacity of PU2 is large enough, while that of PU1 is limited. When
PU1's capacity increases from 4MHz to 24MHz, the prices and the
revenues at the static \emph{NE}s are shown in
Fig.\ref{figure:pricevspu1c} and \ref{figure:revenuevspu1c}. With
the increase of PU1's capacity, both of them tend to reduce the
prices to compete for secondary users. Although the price of PU1
descends, its revenue increases on the contrary due to the increased
capacity. In term of PU2, its price and spectrum demand decrease
until the corresponding values in \emph{Case I} are met. Next, we
analyze the prices and the revenues at the \emph{NE}s when the
capacity of PU2 is constrained by 15MHz and that of PU1 increases
from 4MHz to 24MHz. Fig.\ref{figure:pricevsp1_p2limited} shows that
the prices of PU1 and PU2 decrease when $q_1^a$ grows. In
Fig.\ref{figure:revenuevsp1_p2limited}, the revenue of PU2 becomes
smaller and smaller because the price $p_2^{\dagger}$ decreases
while the spectrum demand is constrained by its capacity.

\begin{figure}[htb]
    \centering
    \includegraphics[width=3.5in]{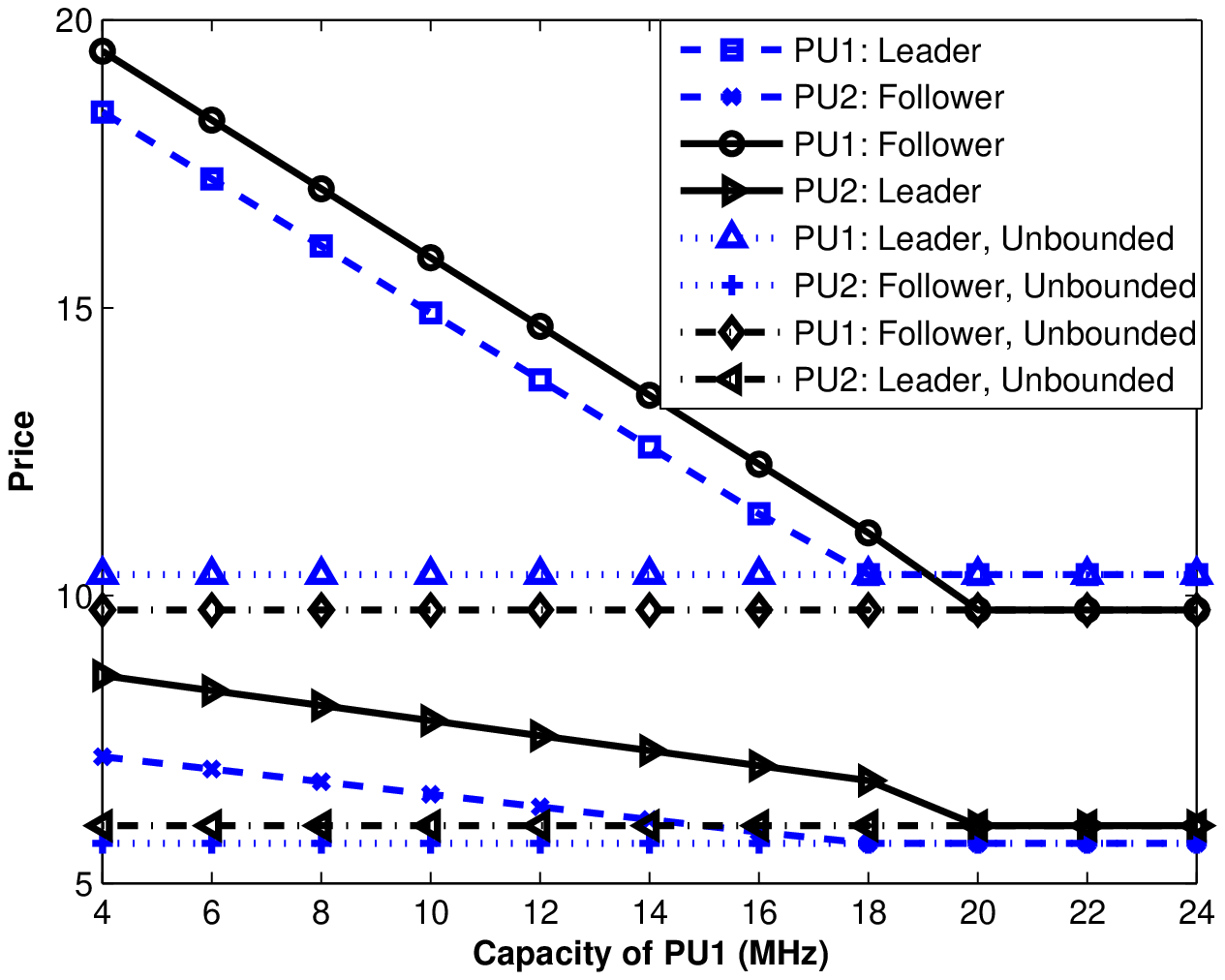}
    \caption{Prices verses PU1's Capacity in the Leader-Follower
    Game for Sufficiently Large $q_2^a$}
    \label{figure:stackelbergpricevspu1c}
\end{figure}

\begin{figure}[!htb]
    \centering
    \includegraphics[width=3.5in]{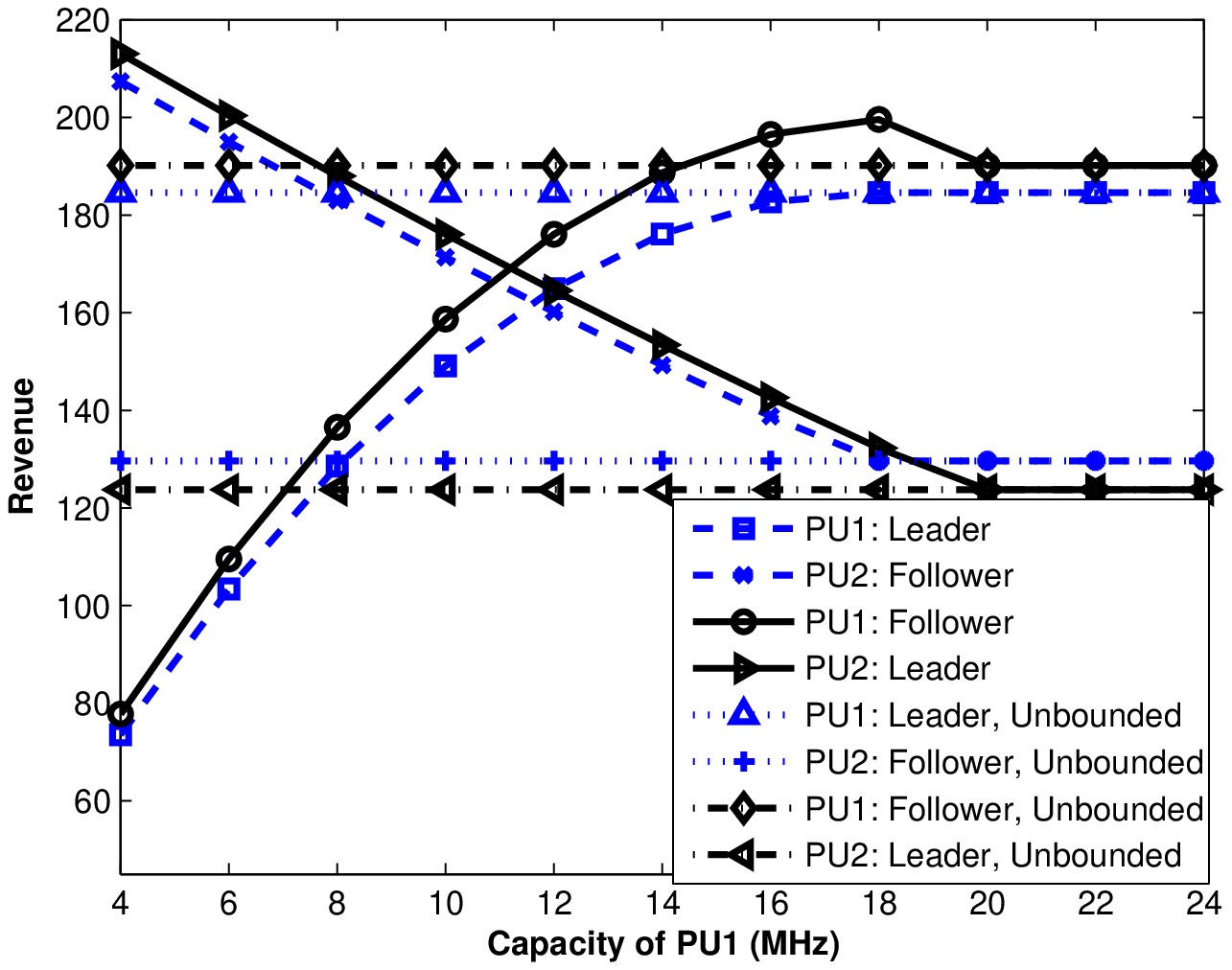}
    \caption{Revenues verses PU1's Capacity in the Leader-Follower
    Game for Sufficiently Large $q_2^a$}
    \label{figure:stackelbergrevenuevspu1c}
\end{figure}

We also illustrate the \emph{NE}s of the leader-follower games. Let
us first consider the case that both primary users have unlimited
capacities. When PU1 is the leader and PU2 is the follower, the
prices at the \emph{NE} are $p_1^{*} = 10.36$ and $p_2^{*} = 5.69$.
The corresponding spectrum demands are $q_1^{*}=17.81$ and
$q_2^{*}=22.77$. When PU2 is the leader and PU1 is the follower, the
prices and the spectrum demands at the \emph{NE} are obtained:
$p_1^{*} = 9.75$, $p_2^{*} = 6$, $q_1^{*}=19.50$ and
$q_2^{*}=20.63$. Next, we study the \emph{NE}s in the case that PU2
has sufficient spectrum and PU1's capacity increases from 4MHz to
24MHz. The prices and revenues are compared in
Fig.\ref{figure:stackelbergpricevspu1c} and
\ref{figure:stackelbergrevenuevspu1c} depending on which primary
user is the leader. When the capacity of PU1 is less than 18MHz, the
primary users have better prices and revenues if PU2 plays the role
of market leader. One can easily draw a conclusion that the primary
user with sufficient capacity, instead of the capacity-insufficient
one, is profitable to be the leader in the duopoly Bertrand game.
When we further increase the PU1's capacity, the \emph{NE}s become
those in \emph{Case 1}.

\subsection{Static Games with Type-II Constraints}

In this subsection, we simulate the competitive pricing of the
static spectrum game with type-II constraints. Specifically, we
evaluate the price of per-unit spectrum and utilities by varying the
capacity constraints and the QoS coefficient $\theta$. Note that the
utility of primary users with type-II constraints contains $\theta
\log (B_i/k_i^{(p)})$ for $i=1,2.$ Since they are constants, we only
compare the parts in the utility functions that are related to the
prices.

\begin{figure}[htb]
    \centering
    \includegraphics[width=3.5in]{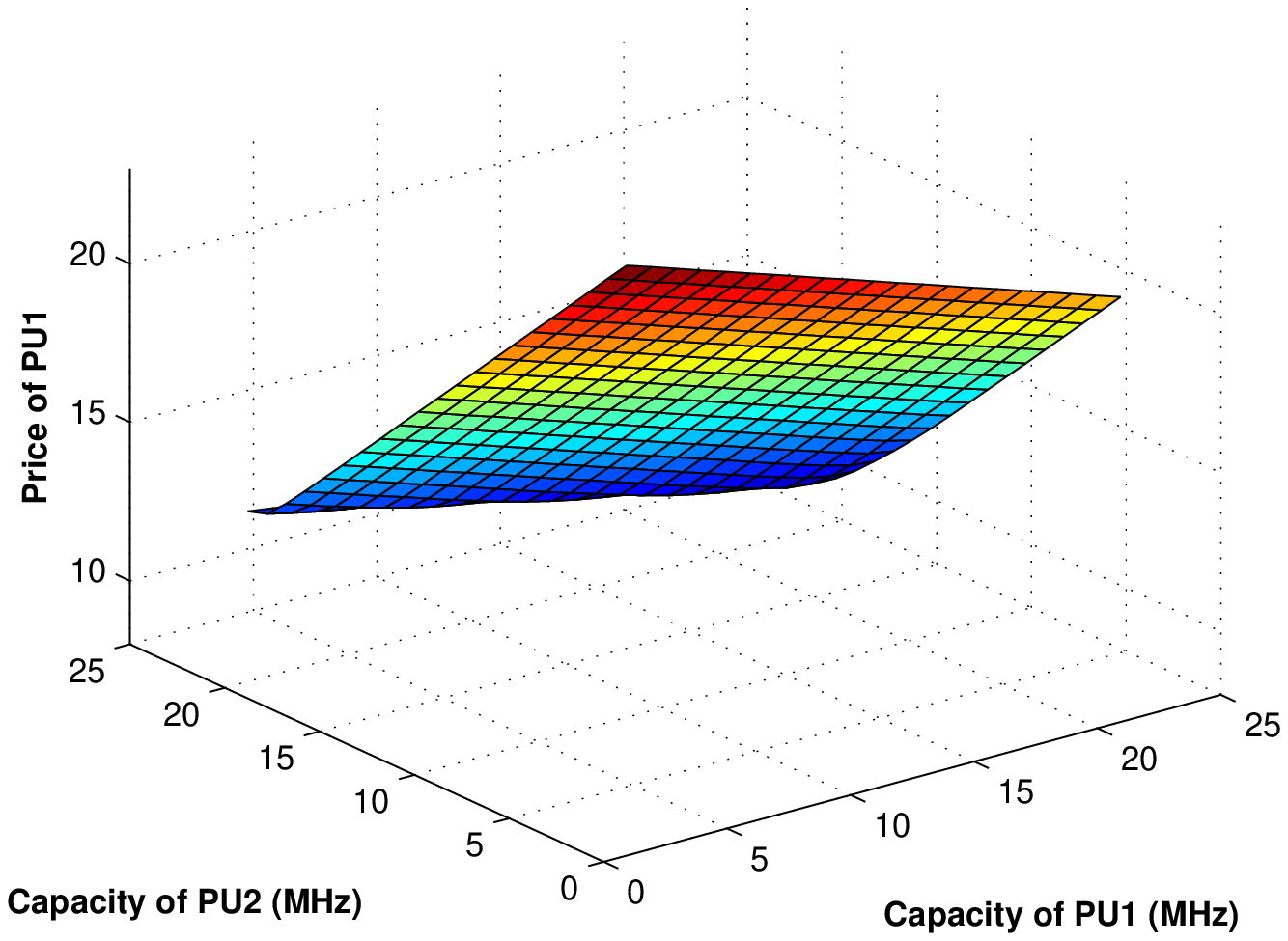}
    \caption{The price of PU1 in type-II model when both primary users increase spectrum capacities}
    \label{figure:type2pu1price}
\end{figure}

\begin{figure}[!htb]
    \centering
    \includegraphics[width=3.5in]{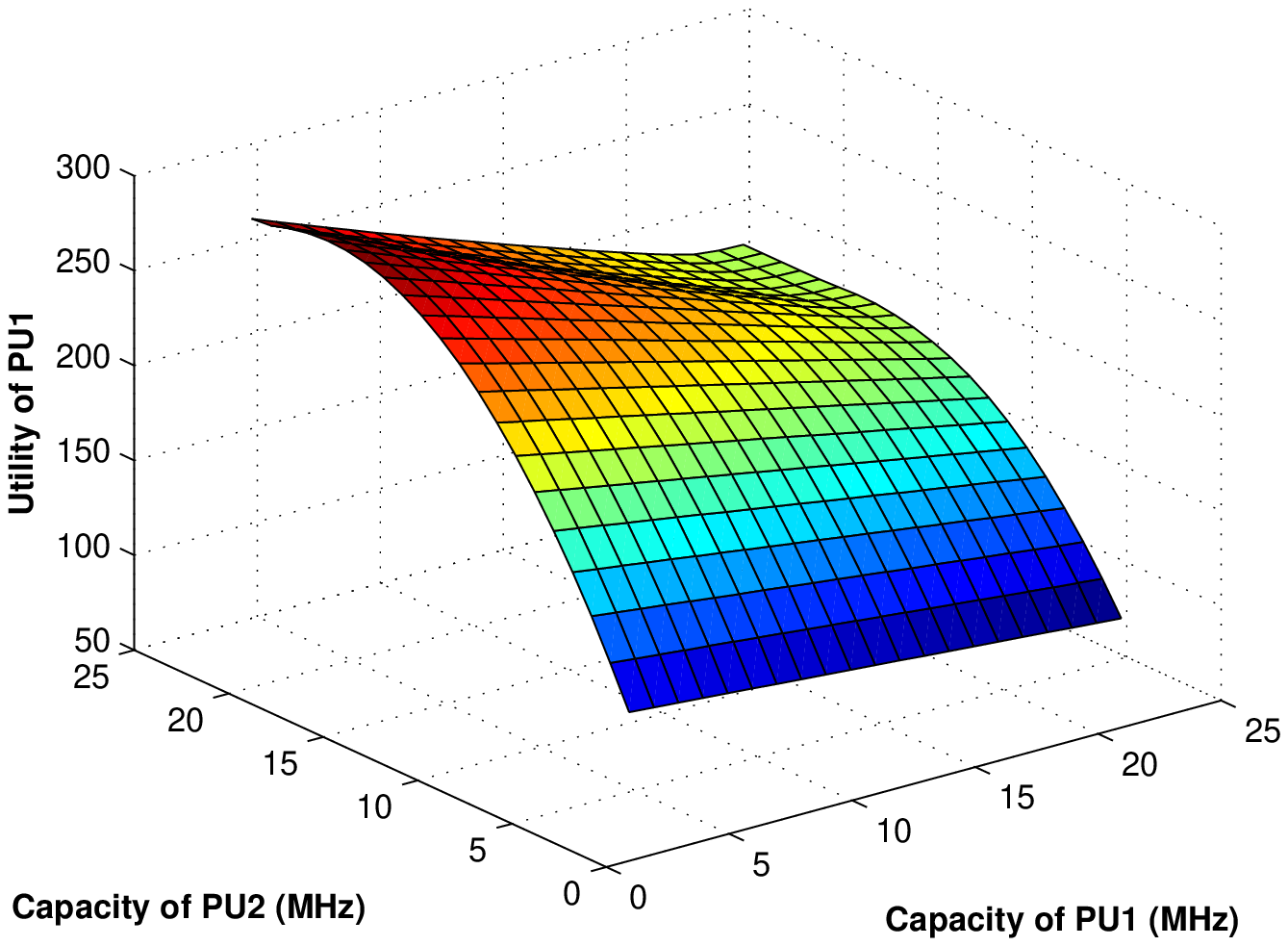}
    \caption{The revenue of PU1 in type-II model when both primary users increase spectrum capacities}
    \label{figure:type2pu1revenue}
\end{figure}

\begin{figure}[!htb]
    \centering
    \includegraphics[width=3.5in]{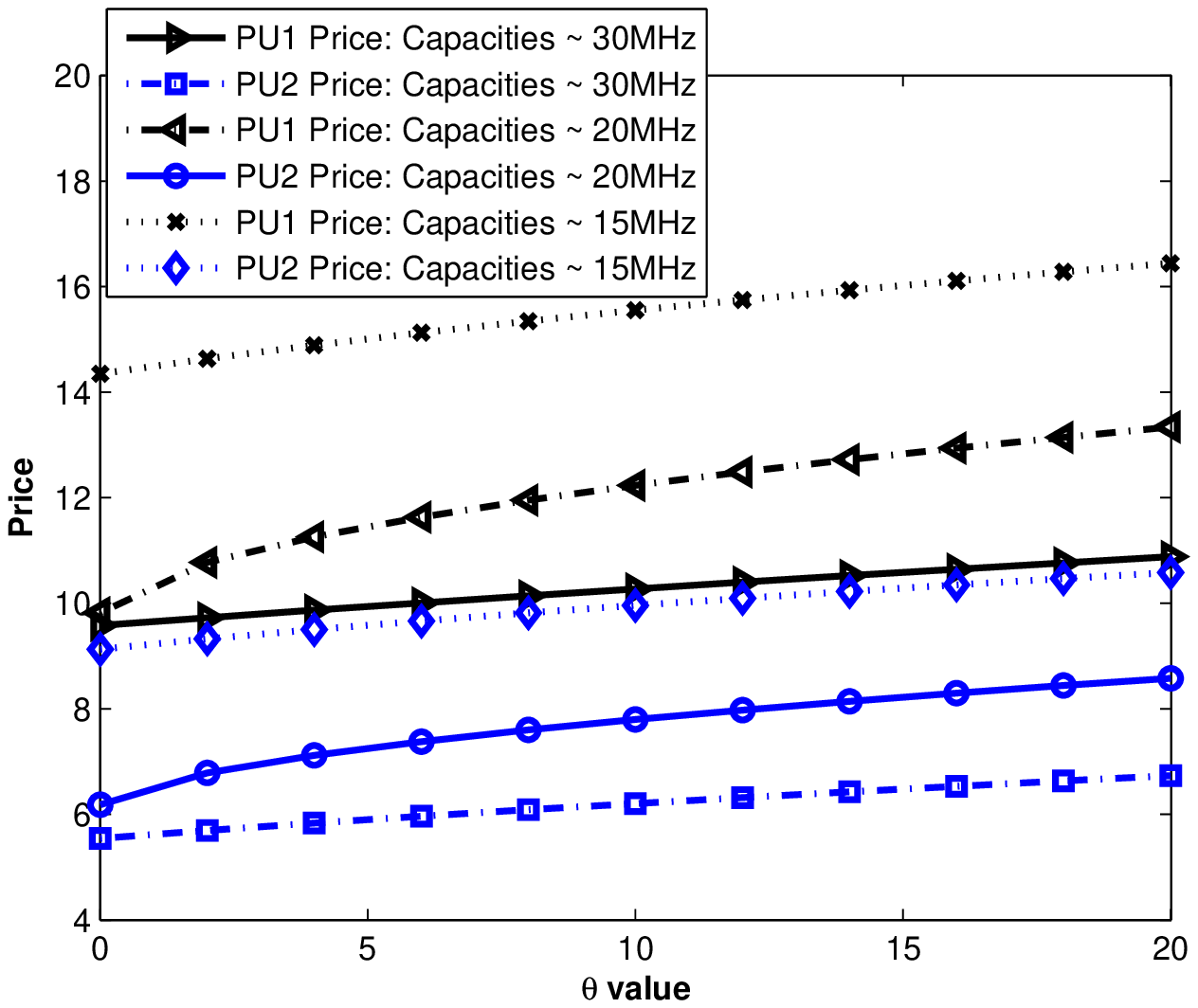}
    \caption{The price of PUs in type-II model when the parameter $\theta$ varies}
    \label{figure:type2thetaprice}
\end{figure}

In the first set of experiments, the coefficient $\theta$ is set to
0.1, and the capacities of primary users increase from 4MHz to
24MHz. The prices of PU1 at the $NE$s are shown in
Fig.\ref{figure:type2pu1price}. The numerical experiments manifest
that the prices of per-unit spectrum decreases with the increase of
the primary users' capacities. Fig.\ref{figure:type2pu1revenue}
demonstrates the utilities of PU1, in which the utility is an
increasing function with respect to the capacity of PU2. One can see
that the utilities grow when PU1 and PU2 increase their capacities.
In the second set of experiments, we aim to explore the relationship
between $\theta$ and the price competition. Let $q_1^a$ and $q_2^a$
be 15MHz. When $\theta$ increases from 0.001 to 20, the prices of
the primary users are shown in Fig.\ref{figure:type2thetaprice}.
When $\theta$ becomes larger, the primary users are inclined to
increase the prices to reduce the utility loss caused by the penalty
functions.


\subsection{Dynamic Game}

\begin{figure}[htb]
  \begin{minipage}[t]{0.99\linewidth}
    \centering
    \includegraphics[width=3.5in]{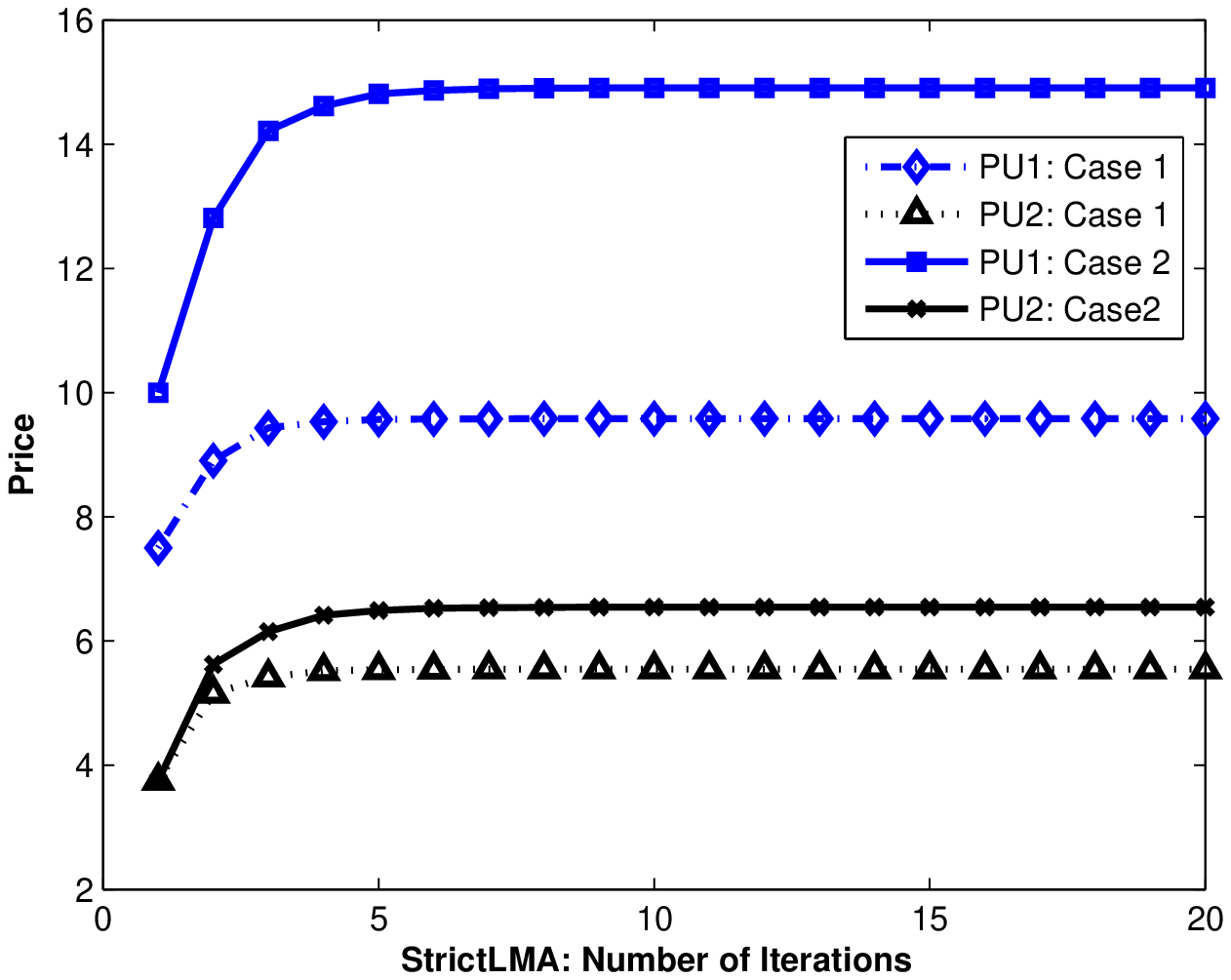}
    \caption{StrictBEST: Iteration of Prices}
    \label{figure:strictlmaprice}
  \end{minipage}
  \begin{minipage}[t]{0.99\linewidth}
    \centering
    \includegraphics[width=3.5in]{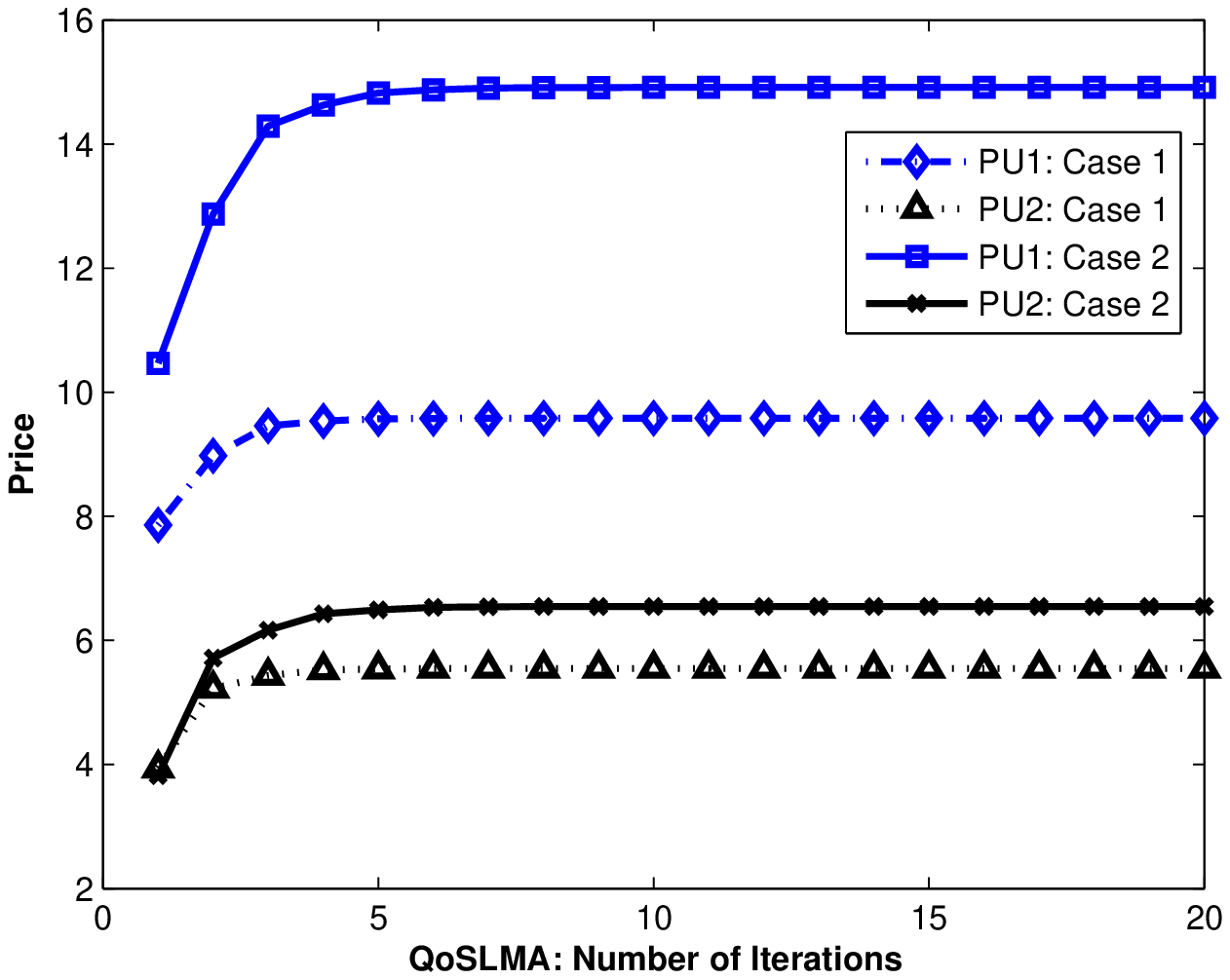}
    \caption{QoSBEST: Iteration of Prices}
    \label{figure:qoslmaprice}
  \end{minipage}
\end{figure}

\begin{figure}[!htb]
  \begin{minipage}[t]{0.99\linewidth}
    \centering
    \includegraphics[width=3.5in]{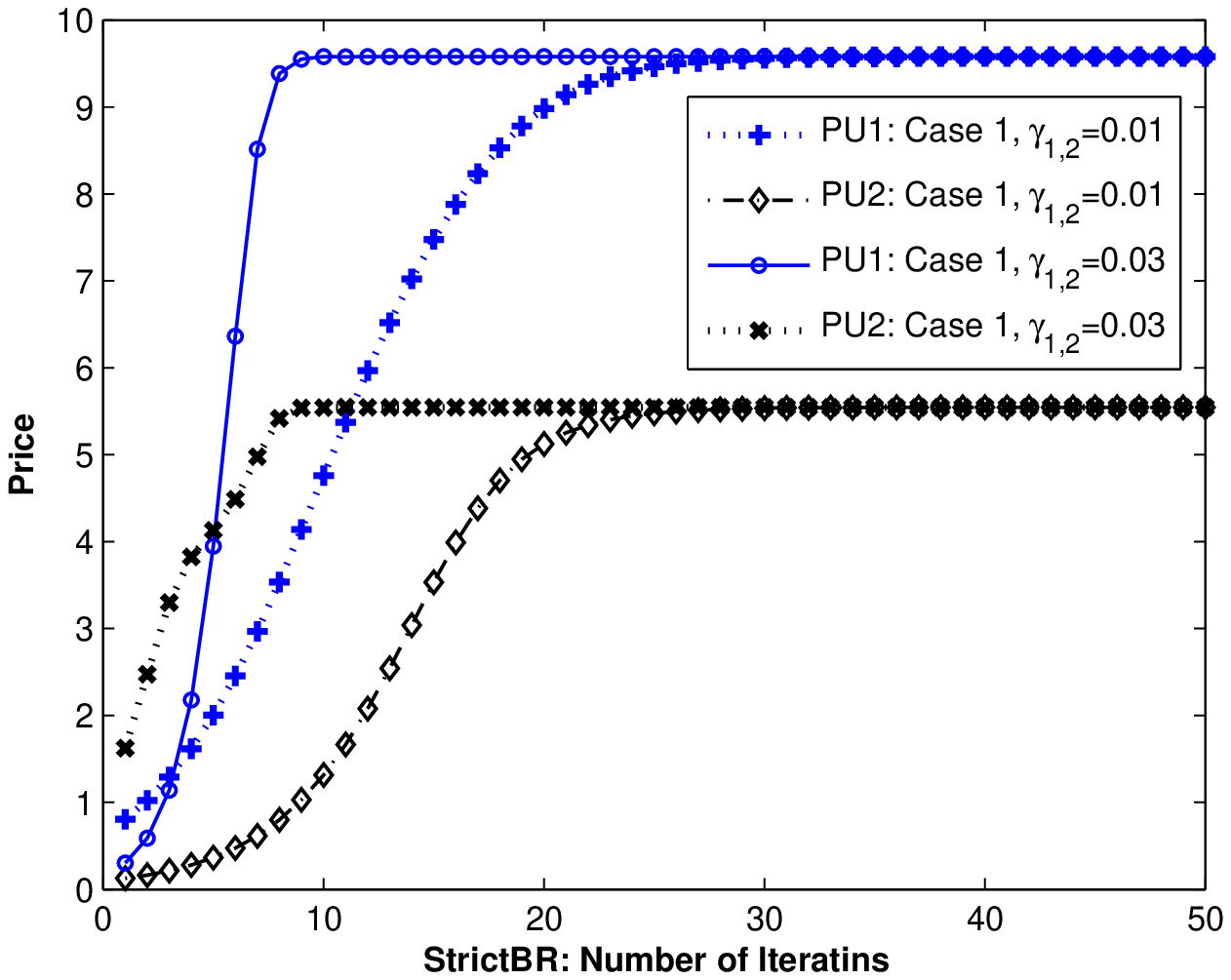}
    \caption{StrictBR: Iteration of Prices: $(q_1^a=100MHz,q_2^a=100MHz)$ }
    \label{figure:strictbrpricecase1}
  \end{minipage}
  \begin{minipage}[t]{0.99\linewidth}
    \centering
    \includegraphics[width=3.5in]{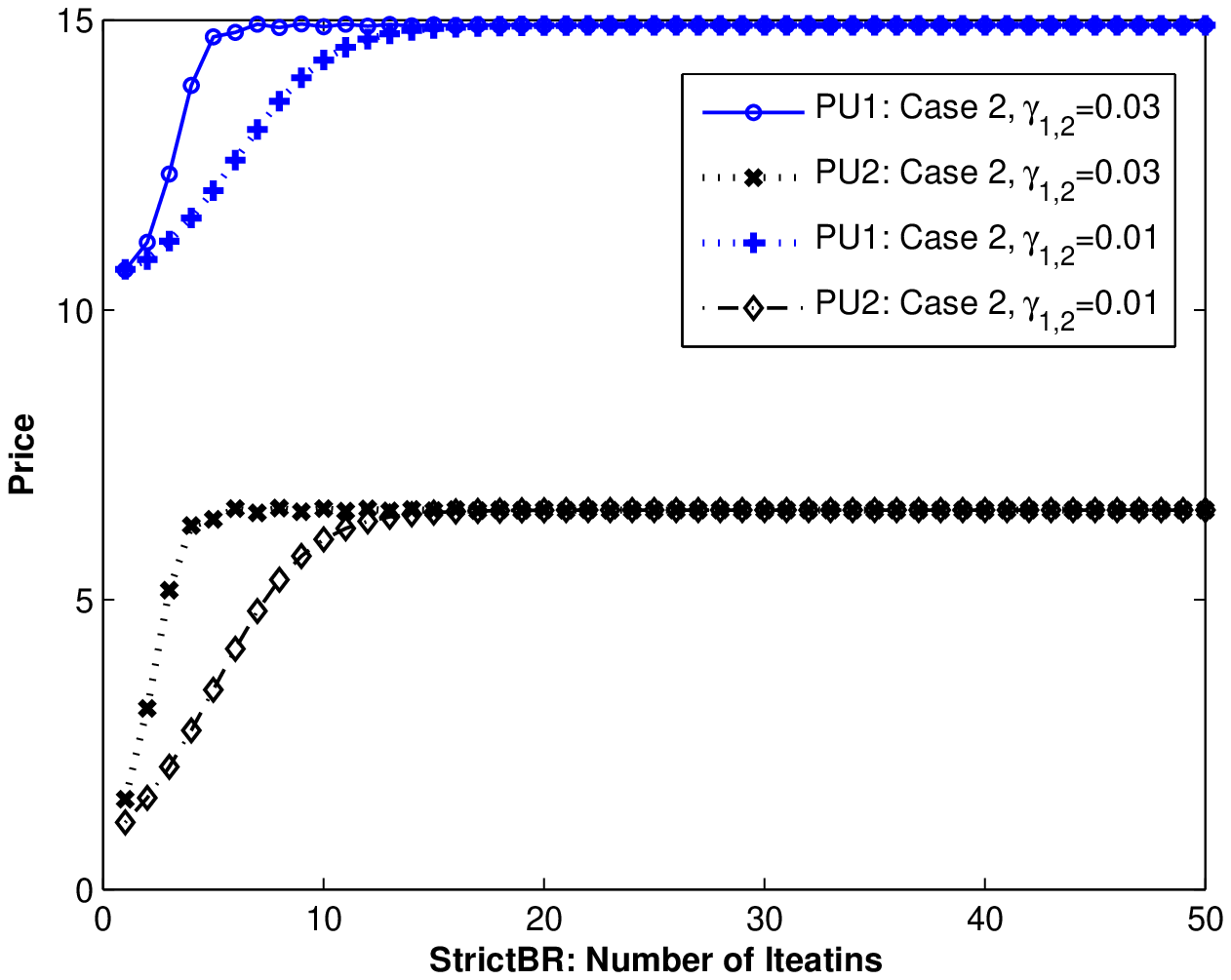}
    \caption{StrictBR: Iteration of Prices: $(q_1^a=10MHz,q_2^a=100MHz)$}
    \label{figure:strictbrpricecase2}
  \end{minipage}
\end{figure}

We examine the dynamic behaviors of the noncooperative games with
\emph{Type-I} and \emph{Type-II} constraints. Especially, the
convergence rates of the proposed algorithms: \emph{StrictBEST},
\emph{StrictBR} and \emph{QoSBEST} are evaluated. In the type-I
model, we evaluate two settings that correspond to \emph{Case 1} and
\emph{Case 2} respectively: $\{q_1^a=100MHz,q_2^a=100MHz\}$ and
$\{q_1^a=10MHz,q_2^a=100MHz\}$. The QoS coefficient $\theta$ is set
to 0.1 in the \emph{QoSBEST} scheme. The price adaptations of
\emph{StrictBEST} and \emph{QoSBEST} are shown in
Fig.\ref{figure:strictlmaprice}-\ref{figure:qoslmaprice}. One can
see that both \emph{StrictBEST} and \emph{QoSBEST} quickly converge
to their individual equilibrium points. The dynamic adjustment of
the \emph{StrictBR} scheme is shown in
Fig.\ref{figure:strictbrpricecase1}-\ref{figure:strictbrpricecase2}
where the learning rates $\gamma_{1,2}$ are both set to 0.01 and
0.03. The convergence rate of \emph{StrictBR} depends on the
learning rates $\gamma_1$ and $\gamma_2$. By cross-comparing
Fig.\ref{figure:strictbrpricecase1} and
Fig.\ref{figure:strictbrpricecase2}, we observe that the convergence
rate of \emph{Case 2} is faster than that of \emph{Case 1}. This is
because the adjustment strategy of PU1 does not have a conservative
learning procedure in \emph{Case 2}. In the \emph{StrictBR} scheme,
small learning rates can guarantee stability of the self-mapping
system, however, at the cost of slow convergence speeds.


\subsection{Nonlinear Instability with Bounded Rationality}

We explore the nonlinear dynamics such as bifurcation and chaos in
the type-I duopoly game with bounded rationality. These complex
behaviors are important because they reveal how prices of primary
users evolve over time and how initial conditions influence the
results of spectrum allocation. As is shown above, the
\emph{StrictBR} scheme can be applied in \emph{Case 1,2,3}. Thus, we
only consider \emph{Case 1} and \emph{Case 2} in the numerical
studies since \emph{Case 3} is similar to \emph{Case 2}.

\begin{figure}[!htb]
  \begin{minipage}[t]{0.99\linewidth}
    \centering
    \includegraphics[width=3.5in]{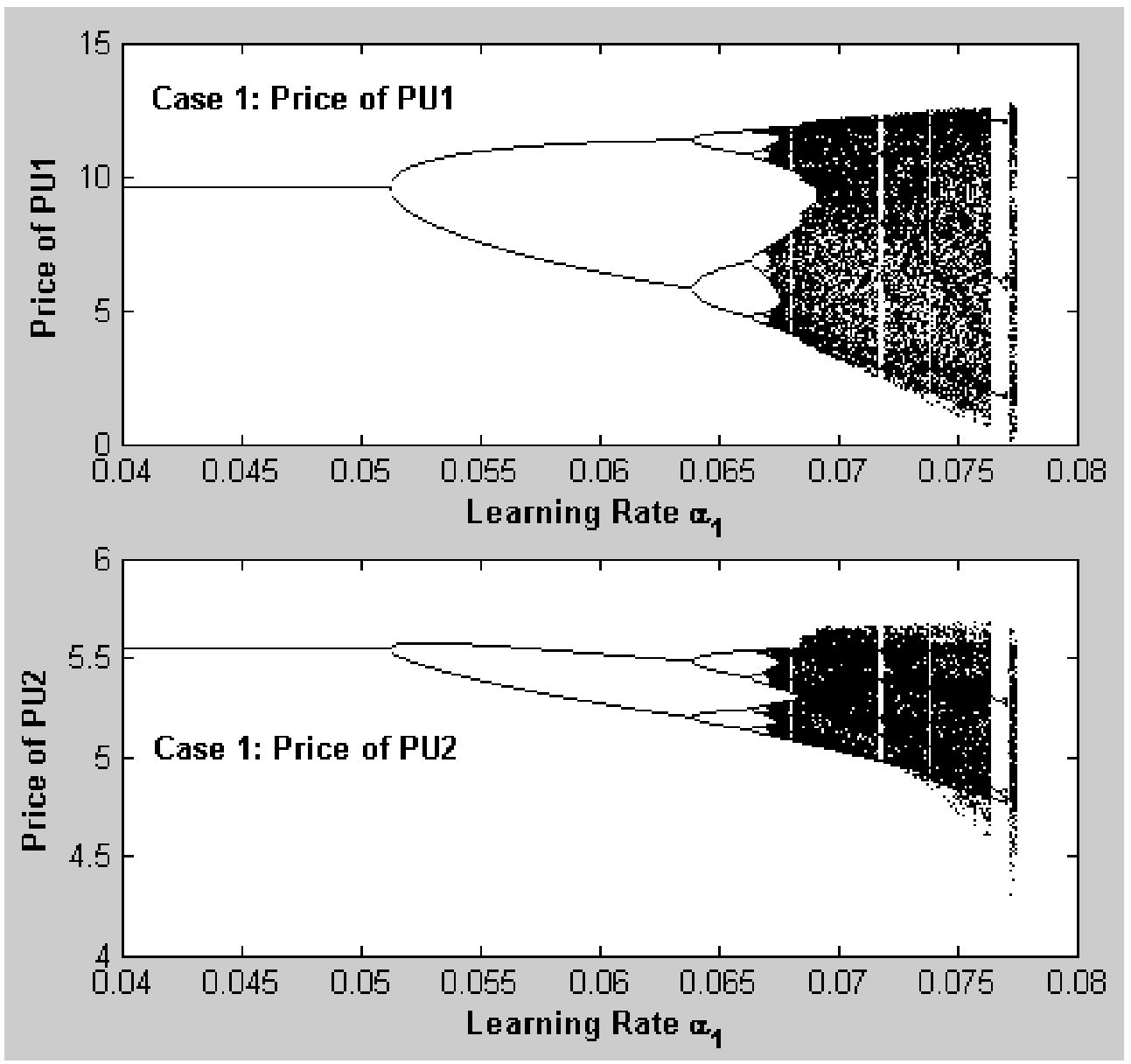}
    \caption{Bifurcation Diagram: $q_1^a = 100MHz$ and $q_2^a = 100MHz$}
    \label{figure:bifurcase1}
  \end{minipage}
  \begin{minipage}[t]{0.99\linewidth}
    \centering
    \includegraphics[width=3.5in, height=3in]{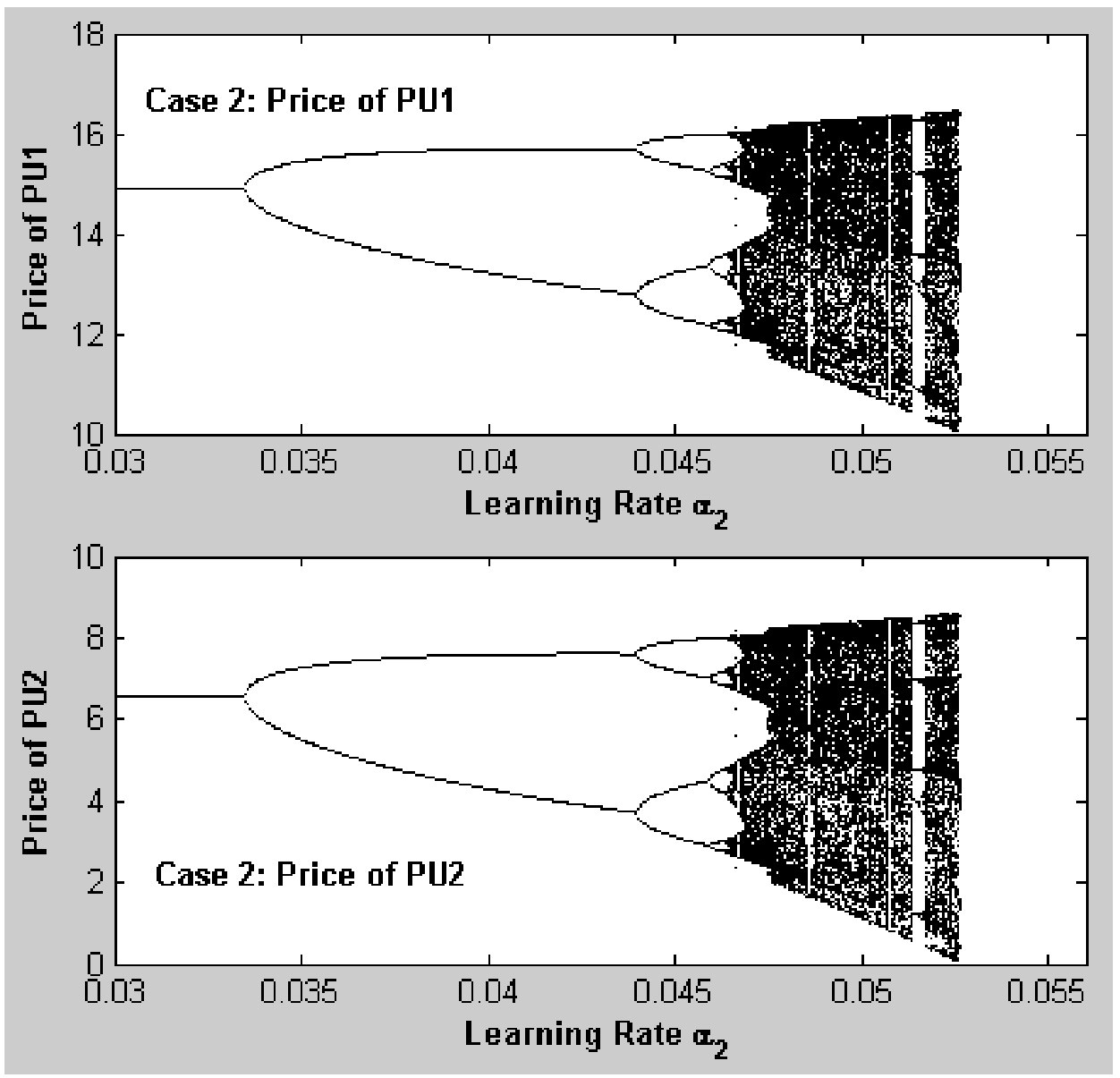}
    \caption{Bifurcation Diagram: $q_1^a = 10MHz$ and $q_2^a = 100MHz$}
    \label{figure:bifurcase2}
  \end{minipage}
\end{figure}

Fig.\ref{figure:bifurcase1} shows the bifurcation diagram of
\emph{Case 1} with respect to the learning rate $\gamma_1$. Here,
the capacities of PU1 and PU2 are both 100MHz. The learning rate
$\gamma_2$ is fixed to be 0.01 and the learning rate $\gamma_1$
increases from 0.01 to 0.09. The bifurcation diagram manifests that
the attractor of \emph{Case 1} model is multivalued in term of
parameter $\gamma_1$. One can also see in
Fig.\ref{figure:bifurcase1} that there exists a stable \emph{NE}
when $\gamma_1$ is less than 0.0511. As $\gamma_1$ further
increases, the \emph{NE} become unstable and infinitely periodic
doubling that leads to chaos eventually. The bifurcation diagram of
\emph{Case 2} with respect to $\gamma_2$ is illustrated in
Fig.\ref{figure:bifurcase2} where the capacity bounds are
$q_1^a=$10MHz and $q_2^a=$100MHz. The learning rate $\gamma_1$ is
0.01 and the learning rate $\gamma_2$ grows from 0.01 to 0.06. When
$\gamma_2$ is less than 0.0331, the \emph{StrictBR} scheme converges
to the unique \emph{NE} of \emph{Case 2} duopoly model.

\begin{figure}[!htb]
  \begin{minipage}[t]{0.99\linewidth}
    \centering
    \includegraphics[width=3.5in]{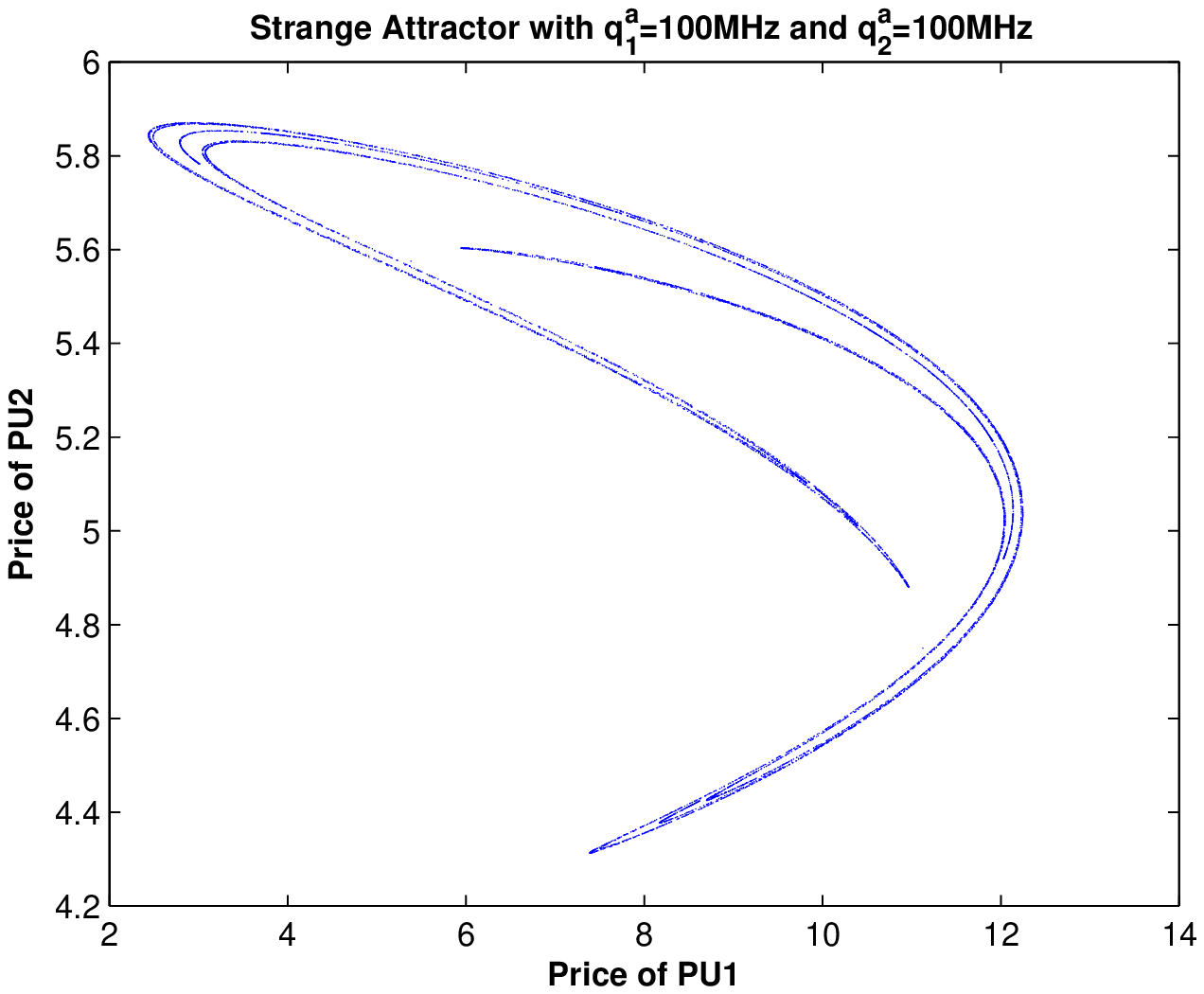}
    \caption{A Strange Attractor for the Following Parameters: Initial
    $p_1=5$, Initial $p_2=5$, $\gamma_1=0.07$, $\gamma_2=0.02$,
    $q_1^a = 100MHz$ and $q_2^a = 100MHz$}
    \label{figure:strangeattractorcase1}
  \end{minipage}
  \begin{minipage}[t]{0.99\linewidth}
    \centering
    \includegraphics[width=3.5in]{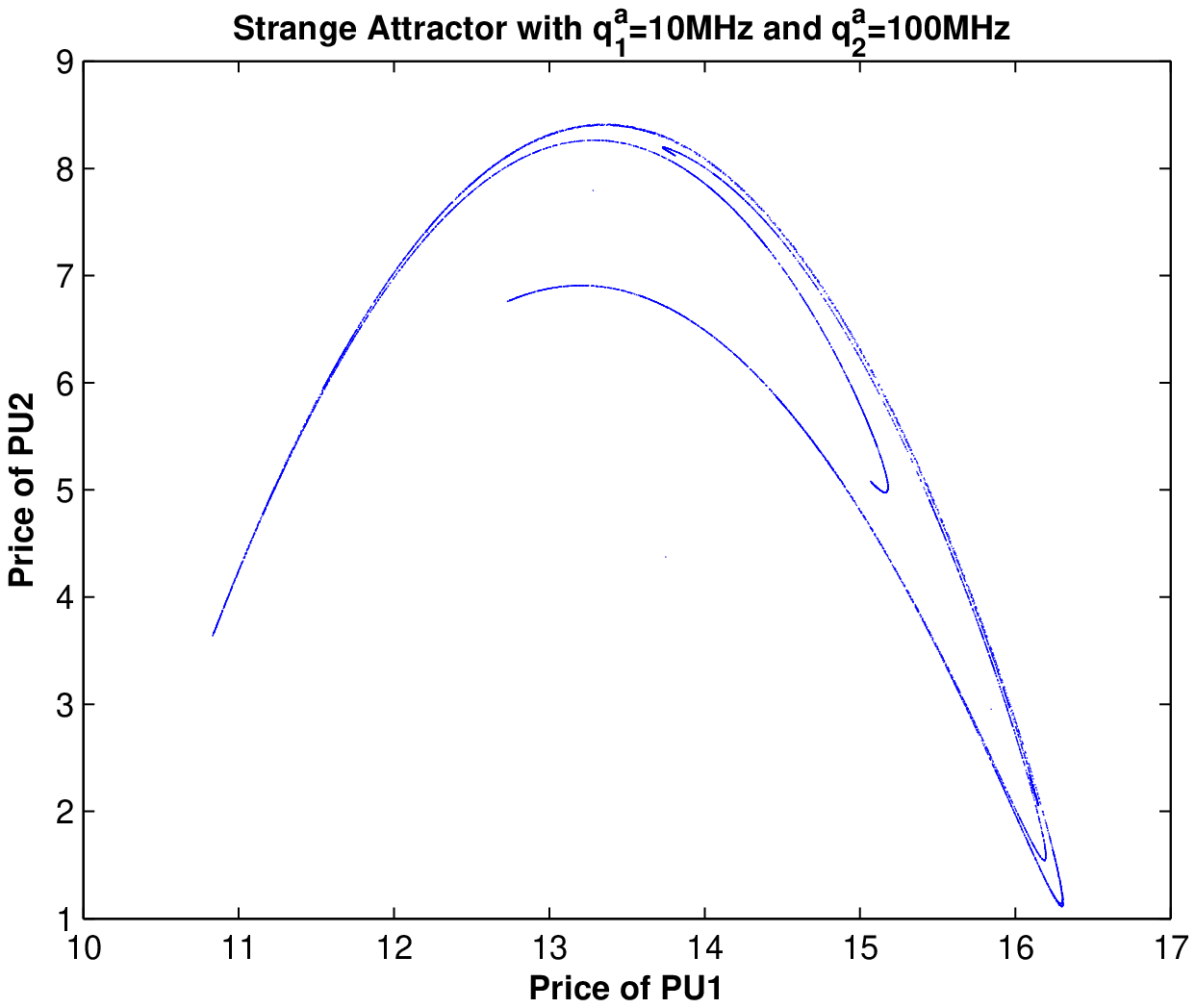}
    \caption{A Strange Attractor for the Following Parameters: Initial
    $p_1=5$, Initial $p_2=5$, $\gamma_1=0.01$, $\gamma_2=0.06$,
    $q_1^a = 10MHz$ and $q_2^a = 100MHz$}
    \label{figure:strangeattractorcase2}
  \end{minipage}
\end{figure}

We show the graphs of strange attractors for \emph{Case 1} with the
parameter constellation $(\gamma_1,\gamma_2)=(0.07, 0.02)$ in
Fig.\ref{figure:strangeattractorcase1} and for \emph{Case 2} with
the parameter constellation $(\gamma_1,\gamma_2)=(0.01, 0.06)$ in
Fig.\ref{figure:strangeattractorcase2}. Especially,
Fig.\ref{figure:strangeattractorcase2} exhibits a fractal structure
similar to Henon attractor \cite{Henon76:Henon}.

\begin{figure}[!htb]
  \begin{minipage}[t]{0.99\linewidth}
    \centering
    \includegraphics[width=3.5in]{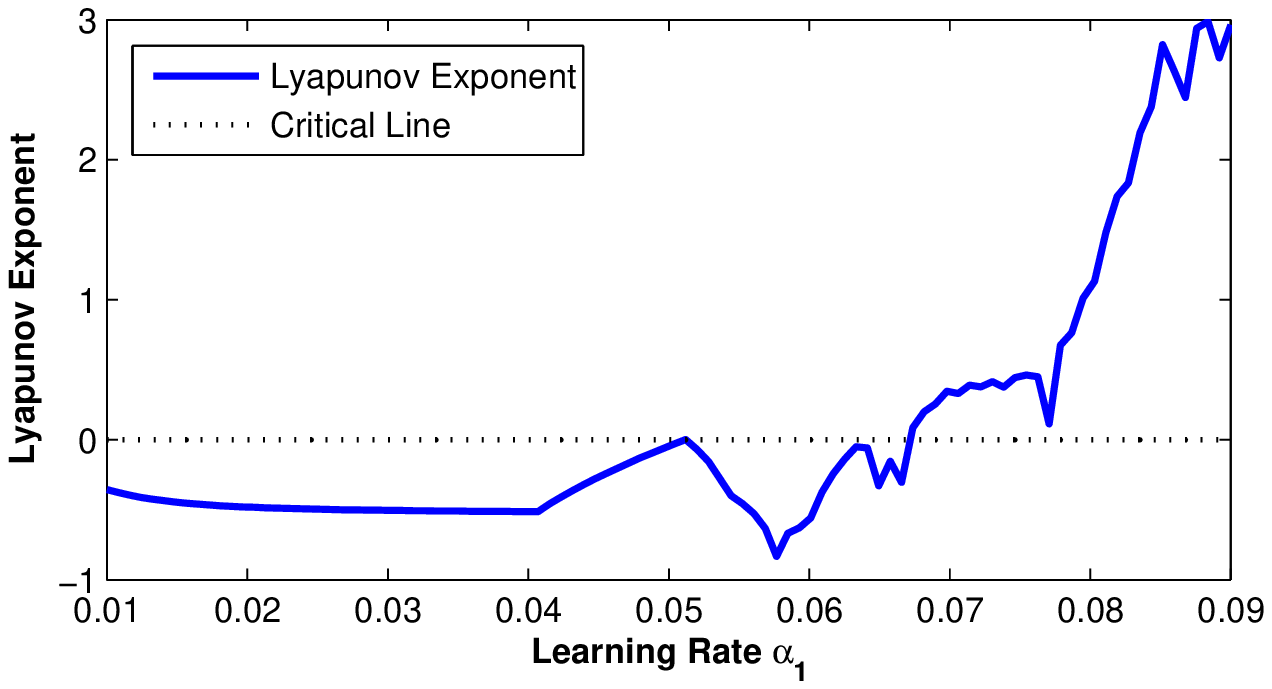}
    \caption{Lyapunov Exponent: $q_1^a = 100MHz$ and $q_2^a = 100MHz$}
    \label{figure:lyaexpcase1}
  \end{minipage}
  \begin{minipage}[t]{0.99\linewidth}
    \centering
    \includegraphics[width=3.5in]{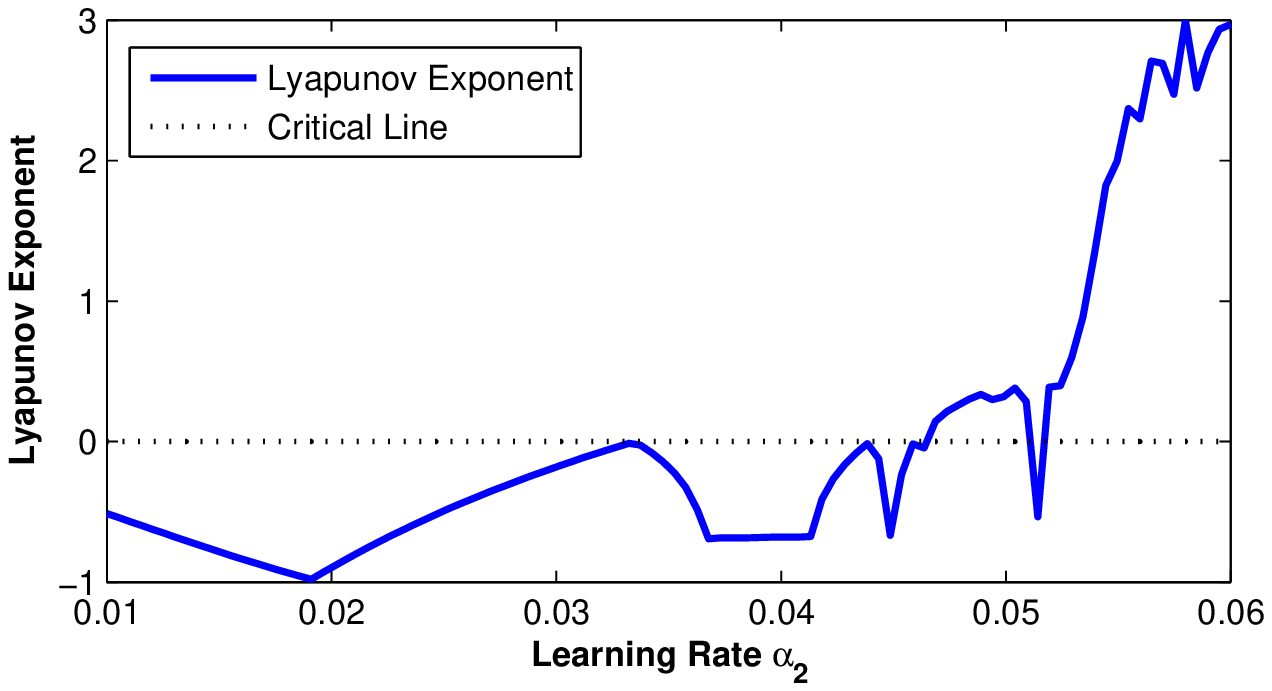}
    \caption{Lyapunov Exponent: $q_1^a = 10MHz$ and $q_2^a = 100MHz$}
    \label{figure:lyaexpcase2}
  \end{minipage}
\end{figure}

The Lyapunov exponent of a dynamical system characterizes the rate
of separation of infinitesimally close trajectories. To analyze the
parameter settings in which aperiodic behaviors occur, we compute
the maximal Lyapunov exponents for the learning rates. If the
maximal Lyapunov exponent is positive, the duopoly game with bounded
rationality is chaotic. For \emph{Case 1}, the maximal Lyapunov
exponent is shown in Fig.\ref{figure:lyaexpcase1} as a function of
the learning rate $\gamma_1$. When $\gamma_1$ is 0.0511, the maximal
Lyapunov exponent becomes positive, which causes the first periodic
doubling bifurcation in Fig.\ref{figure:bifurcase1}. When $\gamma_1$
is greater than 0.0671, the maximal Lyapunov exponent is greater
than 0. This indicates that the self-mapping price adaptation is a
chaotic system. In Fig.\ref{figure:lyaexpcase2}, we display the
maximal Lyapunov exponent of \emph{Case 2} with respect to the
learning rate $\gamma_2$. Here, the learning rate $\gamma_1$ is set
to 0.01. When the learning rate $\gamma_2$ is around 0.0331, the
duopoly game in \emph{Case 2} meets the first doubling bifurcation.
With the increase of $\gamma_2$, the dynamic price adaptation
becomes chaotic.


\section{{\bf Related Work}}
\label{section:related}

The rapid development of wireless communication systems in the past
two decades have resulted in the great needs of a finite and scarce
resource: wireless spectrum. On the other hand, existing wireless
devices operate in the fixed frequency bands, which can be very
inefficient in terms of spectrum utilization. The research carried
out by FCC shows that temporal and geographical variations in the
utilization of the assigned spectrum range from 15\% to 85\%
\cite{FCC03:FCC}. As a promising technology, dynamic spectrum access
is brought forward in the design of next generation wireless
communication systems. The under-utilized spectrum bands can be
detected and exploited by the users equipped with cognitive radios.
For the detailed information, interested readers can refer to recent
surveys in \cite{COMNET06:Ian} and \cite{JSAC05:Haykin}.

One key feature of dynamic spectrum access is how the primary users
(or wireless service provides) and the secondary users (or end
users) share the spectrum efficiently and fairly. In particular, the
dynamic spectrum sharing may involve selling and purchasing
processes. Thus, it is natural to study the interactions of network
components for dynamic spectrum sharing from the perspective of
economics. The existing work can be mainly grouped into two classes:
\emph{auction-based}
\cite{Mobicom08:Zheng,JSAC08:Zhu,Dyspan07:Sengupta} and
\emph{price-based} \cite{JSAC08:Niyato,TWC08:Dusit,Mobihoc08:Zhang}.
Authors in the seminal work \cite{Mobicom08:Zheng} target at the
dishonest bidding issues in an eBay-like dynamic spectrum market. A
truthful and computationally efficient auction mechanism is
presented to perform dynamic spectrum allocation. To maximize
revenue and spectrum utilization, authors in \cite{Dyspan07:Gandhi}
propose a real-time spectrum auction framework to distribute
spectrum among a large number wireless users under interference
constraints. Zhu and Liu \cite{JSAC08:Zhu} propose an auction-based
collusion-resistant dynamic spectrum allocation approach to combat
user collusion in cognitive wireless networks. An economic framework
is also presented in \cite{Dyspan07:Sengupta} to model the spectrum
allocation to wireless service providers (WSPs) and the interaction
of of end users with the WSPs, but the competition among WSPs is not
the focus. Leveraging on microeconomics inspired mechanisms, authors
in \cite{Dyspan07:Grand} develop both bargaining and auction based
mechanisms to find the most optimized allocation pattern for a given
area and allocation duration. Some other auction based spectrum
sharing mechanisms can be found in
\cite{Dyspan07:Ileri,Dyspan08:Wu,Dyspan08:Wu,Mobihoc09:Jia,Mobihoc09:Kasbekar}.
In the price-based class, Niyato and Hossain \cite{JSAC08:Niyato}
introduce the oligopoly pricing theory to characterize the
interactions between the primary users and the secondary users. In
the oligopoly spectrum market, a commonly used quadratic utility is
adopted to quantify the spectrum demand of the secondary service,
and each primary user aims to maximize the individual profit. In
another work \cite{TWC08:Dusit}, they consider the dynamic spectrum
sharing among a primary user and multiple secondary users. They
formulate the problem as an oligopoly market competition and use a
noncooperative game to obtain the Nash Equilibrium. Very recently in
an important work \cite{Mobihoc08:Zhang}, Jia and Zhang formulate
the price and the spectrum competitions as a two-stage
non-cooperative game that is inspired by the theoretic analyses of
Cournot and Bertrand games \cite{RES82:Shaked,RJE84:Singh}. In
\cite{Dyspan07:Daoud}, authors consider a primary user employing
CDMA at the physical layer who aims to lease its spectrum within a
certain geographic subregion. \cite{Dyspan08:Ercan} studies a
revenue maximization problem in a Stackelberg game, where spectrum
owner, primary users and secondary users are the players for
opportunistic spectrum access. Besides, authors in
\cite{Dyspan08:Bae} build a game theoretic model to investigate
whether light regulation in the form of etiquette protocols, device
design and bargaining amongst users can avoid the tragedy of common
in unlicensed spectrum. In terms of nonlinear dynamics in the
economics, authors in \cite{PhysA:Agiza} have shown that the bounded
rationality can cause chaotic behaviors in a Cournot duopoly.

\section{{\bf Conclusion}}
\label{section:conclusion}

This paper suggests an economic framework for dynamic spectrum
allocation in the emerging cognitive radio networks. The primary
users serve as the spectrum brokers that lease the excessive
frequency to the secondary users for monetary payoff. We present
oligopoly Bertrand market models to characterize the
capacity-limited spectrum sharing with two types of constraints: the
strict constraints (type-I) and the QoS penalty functions (type-II).
In the type-I oligopoly market, we present a low-complexity scheme
to search the \emph{NE} and prove its uniqueness. Especially, when
the number of primary users reduces to two, we demonstrate the
interesting revenue gaps in the leader-follower game. Two iterative
algorithm, \emph{StrictBEST} and \emph{StrictBR}, are presented to
adjust the prices when the primary users only possesses the local
market information. In the type-II model, we prove the existence of
unique \emph{NE} and propose a price updating algorithm named
\emph{QoSBEST}. Numerical examples validate our analysis and
manifest the effectiveness of our proposals. In particular, we
experimentally show the representative nonlinear dynamics in the
\emph{StrictBR} algorithm such as bifurcations, chaotic maps as well
as Lyapunov exponents. Our future research will be placed on the
competitive pricing in more complicated markets, e.g. the number of
active primary users are not deterministic.

\section*{{\bf APPENDIX}}
\label{section:appendix}

\noindent \textbf{Lemma \ref{Lemma:PositiveDefinite}:} {\em The
matrix $\mathbf{T}$ is positive definite if the market parameters
has $\beta_i > \mu > 0$ for all $i \in \mathcal{N}$.}

\noindent\textbf{Proof:} Given a $N\times 1$ nonzero, real vector
$\mathbf{x}$, there has,
\begin{eqnarray}
&&\mathbf{x}^{T}\mathbf{T}\mathbf{x} = \mathbf{x}^{T}\left[
\begin{array}{cccc}
\beta_{1} & \mu & \ldots & \mu \\
\mu & \beta_{2} & \ldots & \mu \\
\vdots & \vdots & \ddots & \vdots \\
\mu & \mu & \ldots & \beta_N
\end{array} \right]\mathbf{x} \nonumber \\
&&= \mathbf{x}^{T}\big(\mu\left[
\begin{array}{cccc}
1 & 1 & \ldots & 1 \\
1 & 1 & \ldots & 1 \\
\vdots & \vdots & \ddots & \vdots \\
1 & 1 & \ldots & 1
\end{array} \right] + \left[
\begin{array}{cccc}
\beta_{1} - \mu & 0 & \ldots & 0 \\
0 & \beta_{2} - \mu & \ldots & 0 \\
\vdots & \vdots & \ddots & \vdots \\
0 & 0 & \ldots & \beta_N - \mu
\end{array} \right]\big)\mathbf{x} \nonumber \\
&&= \mu(\sum_{i=1}^{N}x_i)^2 + \sum_{i=1}^{N}(\beta_i-\mu)x_i^2 > 0.
\end{eqnarray}
\noindent Hence, the matrix $\mathbf{T}$ is positive definite if
$\beta_i > \mu > 0$ for all $i \in \mathcal{N}$.\done\\

\noindent \textbf{Lemma \ref{Lemma:PositiveVariables}:} {\em The
parameters that characterize demand-price function in
Eqn.(\ref{equation:sec2.4_demandfunceqn}), i.e.$\; b_i $ and
$c_{ij}$ $(i\neq j)$, are positive, given the conditions $\beta_i >
\mu > 0$ for all $i,j \in \mathcal{N}$.}

\noindent\textbf{Proof:}
We use Cramer's rule to compute the invertible matrix of
$\mathbf{T}$ as follows,

\begin{eqnarray}
\mathbf{T}^{-1}={1 \over
\begin{vmatrix}\mathbf{T}\end{vmatrix}}\left(\mathbf{U}_{ij}\right)^{T}={1
\over
\begin{vmatrix}\mathbf{T}\end{vmatrix}}\left(\mathbf{U}_{ji}\right)={1
\over \begin{vmatrix}\mathbf{T}\end{vmatrix}} \begin{pmatrix}
\mathbf{U}_{11} & \mathbf{U}_{21} & \cdots & \mathbf{U}_{n1} \\
\mathbf{U}_{12} & \mathbf{U}_{22} & \cdots & \mathbf{U}_{n2} \\
\vdots & \vdots & \ddots & \vdots \\ \mathbf{U}_{1n} &
\mathbf{U}_{2n} & \cdots & \mathbf{U}_{nn} \\ \end{pmatrix},
\end{eqnarray}
\noindent where $|\mathbf{T}|$ is the determinant of $\mathbf{T}$
and $\mathbf{U}_{ij}$ is the matrix cofactor. Then, the variable
$b_i$ can be expressed as $b_i = \frac{\mathbf{U}_{ii}}{|T|}$. The
variables $c_{ij}$ can be written as $c_{ij} =
\frac{\mathbf{U}_{ji}}{{|T|}}$ when $i\neq j$. Since $\mathbf{T}$ is
positive definite, $|\mathbf{T}|$ is greater than 0. The cofactor
$\mathbf{U}_{ii}$ is
\begin{eqnarray}
\mathbf{U}_{ii} = {1 \over \begin{vmatrix}\mathbf{T}\end{vmatrix}}
\left|
\begin{array}{cccccc}
\beta_{1} & \ldots & \ldots & \ldots & \ldots & \mu\\
\vdots & \ddots & \ldots & \ldots & \ldots & \mu\\
\vdots & \ldots & \beta_{i-1} & \ldots & \ldots & \mu\\
\vdots & \ldots & \ldots & \beta_{i+1} & \ldots & \mu\\
\vdots & \ldots & \ldots & \ldots & \ddots & \mu\\
\mu & \ldots & \ldots & \ldots & \ldots & \beta_N
\end{array} \right|,\nonumber
\end{eqnarray}
\noindent where all the elements except diagonal ones in the above
determinant are $\mu$. According to Lemma
\ref{Lemma:PositiveDefinite}, one can easily find $b_i >0 \;$ for
all $i\in\mathcal{N}$. Similarly, we can also prove that
$\mathbf{U}_{ji}$ is negative for all $i,j\in\mathcal{N}$ and $i\neq
j$. Thus, the market parameters $c_{ij}$ are all positive for
$i,j\in\mathcal{N}$ and $i\neq j$. The only difference lies in that
we need to exchange certain columns in the determinant before
applying Lemma \ref{Lemma:PositiveDefinite}. \done\\

\noindent \textbf{Lemma \ref{Lemma:Invertible}:} {\em The matrix
$\mathbf{Q}(M_k)$ is positive definite if $\beta_i > \mu > 0$ for
all $i\in\mathcal{N}$ in the utility function.}

\noindent\textbf{Proof:} We prove this lemma by contradiction.

\begin{eqnarray}
\mathbf{Q}(M_k) &=& \left[ \begin{array}{ccccc}
b_{1} & -c_{12} & \ldots & \ldots & -c_{1N} \\
-c_{21} & b_{2} & \ldots & \ldots & -c_{2N} \\
\vdots & \vdots & \ldots & \ldots & \vdots \\
\vdots & \vdots & \ldots & \ldots & \vdots\\
-c_{N1} & -c_{N2} & \ldots & \ldots & b_{N}
\end{array} \right]
+\left[ \begin{array}{ccccc}
0 & \ldots & \ldots & \ldots & \ldots \\
\vdots & \ldots & \ldots & \ldots & \ldots \\
0 & \ldots & b_{M_k+1} & \ldots
& 0 \\
\vdots & \ldots & \ldots & \ldots & \vdots\\
0 & \ldots & \ldots & \ldots & b_{N}
\end{array} \right] \nonumber \\
&=&\left[
\begin{array}{cccc}
\beta_{1} & \mu & \ldots & \mu \\
\mu & \beta_{2} & \ldots & \mu \\
\vdots & \vdots & \ddots & \vdots \\
\mu & \mu & \ldots & \beta_N
\end{array} \right]^{-1} + \left[ \begin{array}{ccccc}
0 & \ldots & \ldots & \ldots & \ldots \\
\vdots & \ldots & \ldots & \ldots & \ldots \\
0 & \ldots & b_{M_k+1} & \ldots
& 0 \\
\vdots & \ldots & \ldots & \ldots & \vdots\\
0 & \ldots & \ldots & \ldots & b_{N}
\end{array} \right]. \nonumber
\end{eqnarray}

We assume that the matrix $Q(M_k)$ is singular. Thus, there exists a
non-zero vector $\mathbf{x}$ that has
$\mathbf{x}^{T}Q(M_k)\mathbf{x} = 0$. Because the matrix
$\mathbf{T}$ is positive definite, it inverse matrix is also
positive definite. We then rewrite $\mathbf{x}^{T}Q(M_k)\mathbf{x}$
as:
\begin{eqnarray}
\mathbf{x}^{T}Q(M_k)\mathbf{x} =
\mathbf{x}^{T}\mathbf{T}^{-1}\mathbf{x} + \sum_{i=M_k+1}^{N}
b_ix_{i}^2 \geq \mathbf{x}^{T}\mathbf{T}^{-1}\mathbf{x} > 0.
\nonumber
\end{eqnarray}
\noindent for any non-zero vector $\mathbf{x}$. \done\\

\noindent \textbf{Lemma \ref{Lemma:Nondecreasing}:} {\em The set of
capacity-insufficient primary users in the $k-1^{th}$ step is a
subset of that in the $k^{th}$ step.}

\noindent\textbf{Proof:} Denote $M_k$ and $\{M_k\}$ to be the number
and the set of capacity-insufficient primary users in the $k^{th}$
search respectively. Denote $p_i^{(k)}$ to be the price of primary
user $i$ in the $k^{th}$ search. At the beginning, $M_0$ is equal to
0.

In the first search, there must have $M_1 > 0$. Otherwise, all the
primary users have sufficient capacities. Without loss of
generality, we look at the the $k^{th}$ search result.

The $k^{th}$ search is based on a priori knowledge that the primary
users in the set $\{M_{k-1}\}$ are capacity-insufficient. We assume
that the $k^{th}$ search finds out $M_{k}$ capacity-insufficient
primary users. Hence, the newly found primary users satisfy
\begin{eqnarray}
\frac{a_i+\sum_{j\neq i}c_{ij}p_j^{(k)}}{2} > q_i^a, \;\;\;\;\;
\forall i=M_{k-1}+1,\cdots,M_{k}.
\end{eqnarray}
\noindent Because the price strategy of the $i^{th}$ new
capacity-insufficient PU is $\frac{a_i+\sum_{j\neq
i}c_{ij}p_j^{(k)}}{2b_i}$, the above equality is equivalent to
$b_ip_i^{(k)} > q_i^a$.

\noindent Next, we compare the price vector of the $k^{th}$ and the
$k-1^{th}$ searches. According to
Eqn.(\ref{equation:sec3.2.5_OligopolyPriceStage2_3}), an alternative
form of price difference is expressed as
\begin{eqnarray}
&& \mathbf{Q}(M_{k})\mathbf{p}^{(k)} -
\mathbf{Q}(M_{k-1})\mathbf{p}^{(k-1)} \nonumber\\
&&= \mathbf{Q}(M_{k-1})\cdot (\mathbf{p}^{(k)} - \mathbf{p}^{(k-1)})
- \left[
\begin{array}{c}
\mathbf{0}\\
b_{M_{k-1}+1}p_{M_{k-1}+1}^{(k)}\\
\vdots\\
b_{M_{k}}p_{M_{k}}^{(k)}\\
\mathbf{0} \end{array} \right] = - \left[
\begin{array}{c}
\mathbf{0}\\
q_{M_{k-1}+1}^a\\
\vdots\\
q_{M_{k}}^a\\
\mathbf{0} \end{array} \right]
\end{eqnarray}
Submit the conditions $b_i^{k}q_i^{(k)} > q_i^a$ for all
$i=M_{k-1}+1,\cdots,M_{k}$ to the above equation, we obtain
\begin{eqnarray}
\mathbf{Q}(M_{k-1})\cdot (\mathbf{p}^{(k)} - \mathbf{p}^{(k-1)}) >
0.
\end{eqnarray}
\noindent Because $\mathbf{Q}(M_{k-1})$ is a Stieltjes matrix, it is
inverse nonnegative. Therefore, all elements in the vector
$\mathbf{p}^{(k)} - \mathbf{p}^{(k-1)}$ are nonnegative. This means
that the prices of all primary users do not decrease in each search.
For the primary users in the set $M_{k-1}$, their optimal spectrum
demands are $\frac{a_i+\sum_{j\neq i}c_{ij}p_j^{(k)}}{2}$, which is
also nondecreasing. To sum up, when a primary user is found to be
capacity-insufficient in the $k-1^{th}$ round, it still lacks of
capacity in the next search.
\done\\

\noindent \textbf{Theorem \ref{Lemma:UniqueNE}:} {\em Consider a
type-I oligopoly spectrum market in
Eqn.(\ref{equation:sec2.4_demandfunceqn}), there exists a unique
Nash Equilibrium.}

\noindent\textbf{Proof:} We prove this theorem via two steps by
contradiction. First, we will show that there exists a unique
\emph{NE} if the capacity-insufficient PUs are unchangeable. In the
second step, we prove that the set of PUs that have insufficient
capacities is unique in the oligopoly market.

As is mentioned earlier, a selfish PU decides the prices according
to the rule Eqn.(\ref{equation:sec3.2.1_OligopolyPrice}) if the
capacity is less than the best demand, and the rule
Eqn.(\ref{equation:sec3.2.2_OligopolyPriceCapaCons}) otherwise.
Provided a market with $N$ PUs, we can find that $M$ of them are
capacity-insufficient through the proposed search method. The price
vector at the \emph{NE}, $\mathbf{p}^{*}$, can be computed by
\begin{eqnarray} \mathbf{p}^{*} = [\mathbf{Q}(M)]^{-1}\cdot \mathbf{a}(M),
\nonumber
\end{eqnarray}
\noindent where $\mathbf{Q}$ is a positive-definite matrix. Hence,
when the set of capacity-insufficient PUs are determined, there is a
unique Nash Equilibrium.

Next, we demonstrate that there has a unique set of
capacity-insufficient primary users. The primary users in the set
$\mathcal{N}$ are grouped into four mutually exclusive classes:
$\mathcal{N}_{1}$, $\mathcal{N}_2$, $\mathcal{N}_3$ and
$\mathcal{N}_4$. PUs in the sets $\mathcal{N}_{1}$ and
$\mathcal{N}_{2}$ are capacity-insufficient in the iterative search
(i.e. $\mathcal{M} = \mathcal{N}_{1}\cup \mathcal{N}_{2}$), but the
PUs in the sets $\mathcal{N}_{3}$ and $\mathcal{N}_{4}$ have enough
capacities. The price vector at the \emph{NE} is denoted as
$\mathbf{p}^{*} = \{p_1^{*},p_2^{*},\cdots,p_N^{*}\}$. At the
\emph{NE}, the PUs must have
\begin{eqnarray}
&&\frac{a_i+\sum_{j\neq i}c_{i,j}p_j^{*}}{2} > q_i^a \;\;
\Rightarrow \;\; b_ip_i^{*} > q_i^a \;\;\; \forall
i\in\mathcal{N}_1; \label{equation:appendix_cond1}\\
&&\frac{a_i+\sum_{j\neq i}c_{i,j}p_j^{*}}{2} > q_i^a \;\;
\Rightarrow \;\; b_ip_i^{*} > q_i^a \;\;\; \forall
i\in\mathcal{N}_2; \label{equation:appendix_cond2}\\
&&\frac{a_i+\sum_{j\neq i}c_{i,j}p_j^{*}}{2} \leq q_i^a \;\;
\Rightarrow \;\; b_ip_i^{*} \leq q_i^a \;\;\; \forall
i\in\mathcal{N}_3; \label{equation:appendix_cond3}\\
&&\frac{a_i+\sum_{j\neq i}c_{i,j}p_j^{*}}{2} \leq q_i^a \;\;
\Rightarrow \;\; b_ip_i^{*} \leq q_i^a \;\;\; \forall
i\in\mathcal{N}_4. \label{equation:appendix_cond4}
\end{eqnarray}

Assume that there exists a different set of PUs that are also
capacity-insufficient. For general purpose, we express this new set
as $\mathcal{N}_1\cup \mathcal{N}_3$ and the capacity-sufficient set
as $\mathcal{N}_2\cup \mathcal{N}_4$. Note that the above PU sets
can be empty, but has a union of $\mathcal{N}$. Since there has a
different set of capacity-insufficient PUs, we can find another
\emph{NE} price vector $\mathbf{p}^{\dag}$ that also have
\begin{eqnarray}
&&\frac{a_i+\sum_{j\neq i}c_{i,j}p_j^{\dag}}{2}
> q_i^a \;\; \Rightarrow \;\; b_ip_i^{\dag} > q_i^a \;\;\; \forall
i\in\mathcal{N}_1; \label{equation:appendix_cond5}\\
&&\frac{a_i+\sum_{j\neq i}c_{i,j}p_j^{\dag}}{2}
> q_i^a \;\; \Rightarrow \;\; b_ip_i^{\dag} > q_i^a \;\;\; \forall
i\in\mathcal{N}_3; \label{equation:appendix_cond6}\\
&&\frac{a_i+\sum_{j\neq i}c_{i,j}p_j^{\dag}}{2} \leq q_i^a \;\;
\Rightarrow \;\; b_ip_i^{\dag} \leq q_i^a \;\;\; \forall
i\in\mathcal{N}_2; \label{equation:appendix_cond7}\\
&&\frac{a_i+\sum_{j\neq i}c_{i,j}p_j^{\dag}}{2} \leq q_i^a \;\;
\Rightarrow \;\; b_ip_i^{\dag} \leq q_i^a \;\;\; \forall
i\in\mathcal{N}_4. \label{equation:appendix_cond8}
\end{eqnarray}
\noindent Here, one can easily find there have $p_i^{\dag}>p_i^{*}$
for $i\in\mathcal{N}_3$ and $p_i^{*}>p_i^{\dag}$ for
$i\in\mathcal{N}_2$. According to the market model, the prices at
the \emph{NE}s can be expressed in terms of spectrum demands
$q_i^{*}$ and $q_i^{\dag}$. Then, $p_i^{\dag}$ and $p_i^{*}$ in the
sets $\mathcal{N}_2$ and $\mathcal{N}_4$ are written by,
\begin{eqnarray}
\!\!\!\!\!&&\!\!\!\!\!\!\!\!\!\!p_i^{*} \!=\! \alpha_i -
(\beta_i-\mu)q_i^a -\mu\sum_{j\in\mathcal{N}_1}q_j^a
-\mu\sum_{j\in\mathcal{N}_2}q_j^a
-\mu\sum_{j\in\mathcal{N}_3}q_j^{*}
-\mu\sum_{j\in\mathcal{N}_4}q_j^{*}, \; \forall i\in
\mathcal{N}_2; \label{equation:appendix_cond9}\\
\!\!\!\!\!&&\!\!\!\!\!\!\!\!\!\!p_i^{*} \!=\! \alpha_i -
(\beta_i-\mu)q_i^{*} -\mu\sum_{j\in\mathcal{N}_1}q_j^a
-\mu\sum_{j\in\mathcal{N}_2}q_j^a
-\mu\sum_{j\in\mathcal{N}_3}q_j^{*}
-\mu\sum_{j\in\mathcal{N}_4}q_j^{*}, \; \forall i\in
\mathcal{N}_3; \label{equation:appendix_cond10}\\
\!\!\!\!\!&&\!\!\!\!\!\!\!\!\!\!p_i^{\dag} \!=\! \alpha_i -
(\beta_i-\mu)q_i^{\dag} -\mu\sum_{j\in\mathcal{N}_1}q_j^a
-\mu\sum_{j\in\mathcal{N}_2}q_j^{\dag}
-\mu\sum_{j\in\mathcal{N}_3}q_j^{a}
-\mu\sum_{j\in\mathcal{N}_4}q_j^{\dag}, \; \forall i\in
\mathcal{N}_2; \label{equation:appendix_cond11}\\
\!\!\!\!\!&&\!\!\!\!\!\!\!\!\!\!p_i^{\dag} \!=\! \alpha_i -
(\beta_i-\mu)q_i^{a} -\mu\sum_{j\in\mathcal{N}_1}q_j^a
-\mu\sum_{j\in\mathcal{N}_2}q_j^{\dag}
-\mu\sum_{j\in\mathcal{N}_3}q_j^{a}
-\mu\sum_{j\in\mathcal{N}_4}q_j^{\dag}, \; \forall i\in
\mathcal{N}_3. \label{equation:appendix_cond12}
\end{eqnarray}
\noindent Recall that the conditions in
Eqn.(\ref{equation:appendix_cond2}),(\ref{equation:appendix_cond3}),(\ref{equation:appendix_cond6})
and (\ref{equation:appendix_cond7}) present the results:
$p_i^{\dag}>p_i^{*}$ for $i\in\mathcal{N}_3$ and
$p_i^{*}>p_i^{\dag}$ for $i\in\mathcal{N}_2$. Hence, we have the
following inequality
\begin{eqnarray}
p_i^{*} + p_l^{\dag} > p_i^{\dag} + p_l^{*} \;\;\; \forall
i\in\mathcal{N}_2, l\in\mathcal{N}_3.
\label{equation:appendix_ineqn1}
\end{eqnarray}
\noindent Submit
Eqn.(\ref{equation:appendix_cond9})-(\ref{equation:appendix_cond12})
to Eqn.(\ref{equation:appendix_ineqn1}) and cancel out the common
items, we obtain the inequality
\begin{eqnarray}
(\beta_i-\mu)q_i^{\dag} + (\beta_l-\mu)q_i^{*} >
(\beta_i-\mu)q_i^{a} + (\beta_l-\mu)q_l^{a} \;\;\; \forall
i\in\mathcal{N}_2, l\in\mathcal{N}_3.
\end{eqnarray}
\noindent Obviously, the above inequality is not true provided the
market conditions $\beta_i
> \mu$ for $i\in \mathcal{N}$. Thus, there is a unique set of
capacity-insufficient primary users, resulting in a unique Nash
Equilibrium in the oligopoly spectrum game. \done\\

\noindent \textbf{Theorem \ref{Theorem:QoSBEST_Converge}:} {\em The
\emph{QoSBEST} algorithm converges to the unique NE if the market
parameters are positive as well as $b_1>c>0$ and $b_2 > c>0$.}

\noindent\textbf{Proof:} Assume the NE prices of PU1 and PU2 are
$p_1^{*}$ and $p_2^{*}$. We start from time $t$ and solve the
equation in Eqn.(\ref{equation:sec4.3.1}). Although $p_1(t+1)$ in
Eqn.(\ref{equation:sec4.3.1}) has two roots, one can easily check
their feasibility by submitting these roots to the equation $q_1(t)
= a_1 - b_1p_1(t+1) +cp_2(t)$. After excluding the infeasible root,
the unique price of PU1 is expressed as
\begin{eqnarray}
p_1(t+1) =
\frac{(3a_1+3cp_2(t)-2q_1^a)+\sqrt{(a_1+cp_2(t)-2q_1^a)^2+8b_1\theta}}{4b_1},
\nonumber
\end{eqnarray}
and the optimal price of PU1 is written as
\begin{eqnarray}
p_1^{*} =
\frac{(3a_1+3cp_2^{*}-2q_1^a)+\sqrt{(a_1+cp_2^{*}-2q_1^a)^2+8b_1\theta}}{4b_1},
\nonumber
\end{eqnarray}

\noindent In order to compare $p_1(t+1)$ and $p_1^{*}$, we consider
four cases with regard to different $p_2(t)$ and $p_2^{*}$.

\noindent\emph{\textbf{Case 1:}} $a_1+cp_2(t)-2q_1^a \geq 0$ and
$a_1+cp_2^{*}-2q_1^a \geq 0$;

\noindent The difference between $p_1(t+1)$ and $p_1^{*}$ is,
\begin{eqnarray}
p_1(t+1) - p_1^{*} = \frac{3c(p_2(t)-p_2^{*})}{4b_1} +
\frac{\sqrt{(a_1+cp_2(t)-2q_1^a)^2+8b_1\theta} -
\sqrt{(a_1+cp_2^{*}-2q_1^a)^2+8b_1\theta}}{4b_1}. \nonumber
\end{eqnarray}
\noindent If $p_2(t) \geq p_2^{*}$, there has $a_1+cp_2(t)-2q_1^a >
a_1+cp_2^{*}-2q_1^a > 0$. According to Lemma \ref{Lemma:Inequality},
the following inequality holds,
\begin{eqnarray}
p_1(t+1) - p_1^{*} \leq \frac{c(p_2(t)-p_2^{*})}{b_1}.
\label{equation:appendix_th4_10}
\end{eqnarray}

\noindent On the other hand, if $p_2(t) < p_2^{*}$, there has
$0<a_1+cp_2(t)-2q_1^a < a_1+cp_2^{*}-2q_1^a$. Based on Lemma
\ref{Lemma:Inequality}, we can obtain the following inequality,
\begin{eqnarray}
p_1(t+1) - p_1^{*} > \frac{c(p_2(t)-p_2^{*})}{b_1}.
\label{equation:appendix_th4_11}
\end{eqnarray}
\noindent Combine Eqn.(\ref{equation:appendix_th4_10}) and
(\ref{equation:appendix_th4_11}) together, we have,
\begin{eqnarray}
|p_1(t+1) - p_1^{*}| \leq \frac{c}{b_1}|p_2(t) - p_2^{*}|.
\label{equation:appendix_th4_12}
\end{eqnarray}

\noindent\emph{\textbf{Case 2:}} $a_1+cp_2(t)-2q_1^a < 0$ and
$a_1+cp_2^{*}-2q_1^a < 0$;

\noindent If $p_2(t) \leq p_2^{*}$, there has $0< 2q_1^a
-(a_1+cp_2^{*}) \leq 2q_1^a - (a_1+cp_2(t))$. Similarly, we
calculate the difference between $p_1(t+1)$ and $p_1^{*}$,
\begin{eqnarray}
\frac{3c(p_2(t)-p_2^{*})}{4b_1} \leq p_1(t+1) - p_1^{*} \leq
\frac{c(p_2(t)-p_2^{*})}{2b_1}. \label{equation:appendix_th4_13}
\end{eqnarray}
\noindent Otherwise, if  $p_2(t) > p_2^{*}$, the difference between
$p_1(t+1)$ and $p_1^{*}$ satisfies,
\begin{eqnarray}
\frac{c(p_2(t)-p_2^{*})}{2b_1} \leq p_1(t+1) - p_1^{*} \leq
\frac{3c(p_2(t)-p_2^{*})}{4b_1}. \label{equation:appendix_th4_13}
\end{eqnarray}
\noindent The above analysis manifests that the following
inequalities hold,
\begin{eqnarray}
\frac{c}{2b_1}|p_2(t)-p_2^{*}| \leq |p_1(t+1) - p_1^{*}| \leq
\frac{3c}{4b_1}|p_2(t)-p_2^{*}|. \label{equation:appendix_th4_13}
\end{eqnarray}

\noindent\emph{\textbf{Case 3:}} $a_1+cp_2(t)-2q_1^a \geq 0$ and
$a_1+cp_2^{*}-2q_1^a < 0$;

\noindent The \emph{case 3} also implies $p_2(t) \geq p_2^{*}$.
According to Lemma \ref{Lemma:Inequality}, the following
inequalities hold,
\begin{eqnarray}
-c(p_2(t)-p_2^{*}) \leq \sqrt{(a_1+cp_2(t)-2q_1^a)^2+8b_1\theta} -
\sqrt{(2q_1^a-(a_1+cp_2^{*}))^2+8b_1\theta} \leq c(p_2(t)-p_2^{*}).
\nonumber
\end{eqnarray}
\noindent Therefore, the difference between $p_1(t+1)$ and $p_1^{*}$
satisfies,
\begin{eqnarray}
\frac{c}{2b_1}|p_2(t)-p_2^{*}| \leq |p_1(t+1) - p_1^{*}| \leq
\frac{c}{b_1}|p_2(t)-p_2^{*}|. \label{equation:appendix_th4_15}
\end{eqnarray}

\noindent\emph{\textbf{Case 4:}} $a_1+cp_2(t)-2q_1^a < 0$ and
$a_1+cp_2^{*}-2q_1^a \geq 0$;

\noindent This case implies $p_2(t) \leq p_2^{*}$. Using the similar
method as that in \emph{Case 3}, we obtain the same inequality in
(\ref{equation:appendix_th4_15}). Combine all the analytic results
together, we can see the distance between $p_1(t+1)$ and $p_1^{*}$
has $|p_1(t+1) - p_1^{*}| \leq \frac{c}{b_1}|p_2(t) - p_2^{*}|$.
Using the similar steps, we can easily find that in slot $(t+1)$,
the following inequality holds,
\begin{eqnarray}
|p_2(t+2) - p_2^{*}| \leq \frac{c}{b_2}|p_1(t+1) - p_1^{*}| \leq
\frac{c^2}{b_1b_2}|p_2(t) - p_2^{*}|\leq
(\frac{c^2}{b_1b_2})^{\frac{t}{2}}|p_2(0) - p_2^{*}|,
\label{equation:appendix_th4_16}
\end{eqnarray}
\noindent where $p_2(0)$ is the initial price of PU2. Given the
conditions $b_1>c$ and $b_2>c$, the \emph{QoSBEST} algorithm
converges to the unique NE.

\end{document}